\documentclass[12pt]{article}
\usepackage{amsmath,amssymb}
\usepackage{graphics}
\usepackage{caption}
\usepackage{subcaption}
\usepackage{latexsym,ifthen}
\usepackage{booktabs}
\usepackage{epsfig,here}
\usepackage{authblk}

\newcommand{\Section}[1]{\section{#1}\setcounter{figure}{0}\setcounter{table}{0}\setcounter{equation}{0}}

\newtheorem{theo}{Theorem}[section]

\title{A thermodynamically consistent phase-field model for two-phase flows with thermocapillary effects}
\author[1,2]{Zhenlin Guo}\author[1]{Ping Lin \thanks{Corresponding author. E-mail: {\tt plin@maths.dundee.ac.uk} }}
\affil[1]{Department of Mathematics, University of Dundee, Dundee DD1
4HN, Scotland, United Kingdom.}
\affil[2]{Department of Applied Mathematics and Mechanics, University of Science and Technology Beijing, Beijing, 100083, China}

\date{}
\begin{document}
\maketitle

\begin{abstract}
In this paper, we develop a phase-field model for binary incompressible (quasi-incompressible) fluid with thermocapillary effects, which allows for the different properties (densities, viscosities and heat conductivities) of each component while maintaining thermodynamic consistency. The governing equations of the model including the Navier-Stokes equations with additional stress term, Cahn-Hilliard equations and energy balance equation are derived within a thermodynamic framework based on entropy generation, which guarantees thermodynamic consistency. A sharp-interface limit analysis is carried out to show that the interfacial conditions of the classical sharp-interface models can be recovered from our phase-field model. Moreover, some numerical examples including thermocapillary convections in a two-layer fluid system and thermocapillary migration of a drop are computed using a continuous finite element method. The results are compared to the corresponding analytical solutions and the existing numerical results as validations for our model.
\end{abstract}
\begin{itemize}
\item[] {\bf Keywords:} Two-phase flows, Phase-field method, Thermocapillary effetcs,\\
\hspace{20mm} ~~~~~~~~Thermodynamic consistency, Entropy generation, Quasi-incompressible. 


\end{itemize}
\Section{Introduction}
When the interface separating two fluids is exposed to a temperature gradient, the variations of surface tension along the interface lead to shear stresses that act on the fluid through viscous forces, and thus induce a motion of the fluids in the direction of the temperature gradient. For most of the fluids, the surface tension generally decreases with the increasing temperature. The non-uniformity of surface tension then drives the fluids to move from the region with higher temperature to that with lower temperature. This effect is known as thermocapillary (Marangoni) effect \cite{Levich1962}, and it plays an important role in various industrial applications involving microgravity \cite{Subramanian2001} or microdevices \cite{Darhuber2005}, where the surface forces become dominant. One famous example for thermocapillary effects is the thermocapillary migration of drops, where the drops are set in a liquid possessing a temperature gradient, and will move toward the hot region due to the thermocapillary effects. The thermocapillary migration of a gas bubble was first examined experimentally by Young $et$ $al.$ \cite{Young1959}, who derived an analytical expression for the terminal velocity of a single spherical drop in a constant temperature gradient by assuming the convective transport of momentum and energy are negligible. Since then, extensive works were carried out experimentally, analytically and numerically in order to investigate this phenomenon, where many of them are summarized by Subramanian and Balasubramaniam \cite{Subramanian2001}. Another example for thermocapillary effects is the thermocapillary convection in a two-layer fluid system (thermocapillary instabilities), where the system is typically confined between two parallel plates and subjected to a temperature gradient. Due to the perturbations in the temperature and velocity field as well as the interface position, surface tension gradients will occur at the interface and drive the fluid to motion. The instabilities then set in and lead to the convective motion, where a typical convection pattern is the hexagonal cell formation found by {{B$\operatorname{\acute{e}}$nard}}  \cite{Benard1900}. The thermocapillary instabilities are widely studied which can be traced back to some pioneering works performed by Block \cite{Block1956}, Pearson \cite{Pearson1958}, and Sternling and Scriven \cite{Sternling1959,Scriven1964}. Literature review of recent experimental and analytical work on instabilities in thermocapillary convection are provided by Schatz and Neitzel \cite{Schatz2001}, Davis \cite{Davis1987} and Andereck $et$ $al.$ \cite{Andereck1998}.\\\\
The problem described above is the multiphase flow problem, where the available numerical methods can roughly be divided into two categories: interface tracking and interface capturing methods. In interface tracking methods, the position of the interface is explicitly tracked, which requires meshes that track the interfaces and are updated as the flow evolves. Boundary integral methods (see the review \cite{Hou2001}), front-tracking methods (see the review \cite{Tryggvason2001, Hua2008}), and immersed boundary methods (see the review \cite{Mittal2005}) are examples of this type. In the context of the multiphase flow with thermocapillary (Marangoni) effects, e.g., the thermocapillary migration and thermocapillary instabilities, several works have been performed by using interface tracking methods. Here we refer \cite{Zhou1996, Berejnov2001, Rother2002} as examples for boundary-integral methods, \cite{Tavener2002, Nas2003, Nas2006, Yin2008} for front-tracking methods, and \cite{Pozrikidis2004, Blyth2004} for immersed-boundary methods. In interface capturing methods, on the other hand, the interface is not tracked explicitly, but instead is implicitly defined through an interface function (e.g. level-set, color or phase-field function). This means that the computations are based on fixed spatial domains and thus eliminate the problem of updating the meshes encountered in interface tracking methods. For example, volume-of-fluid (VOF) methods (see \cite{Scardovelli1999} for the review, and see \cite{Gambaryan-Roisman2005,Ma2013} as examples for thermocapillary effects), level-set methods (see \cite{Osher2001,Sethian2003} for the review, and see \cite{Haj1997, Herrmann2008} as examples for thermocapillary effects) are of this type.\\\\
Another interface capturing method is phase-field method, or diffuse-interface method (see the review \cite{Anderson1998,Emmerich2008,Kim2012}), which has now emerged as a powerful method to simulate many types of multiphase flows, including drop coalescence, break-up, rising and deformations in shear flows \cite{Jacqmin1999,Lee2001,Lee2002, Boyer2002, LiuShen2002, Baldalassi2004, Yue2004, Kim2005, Yue2006, Ding2007, Shen2009, Hua2011}, phase separation \cite{Baldalassi2004,Kim2004,Kim20051}, contact line dynamics \cite{Jacqmin2000, He2011, Gao2012, Bao2012, Jiang2014}, and dynamics of interface with surfactant adsorption \cite{Sman2006, Teigen2011} and thermocapillary effects \cite{Jasnow1996, Borcia2003,Borcia2004,Sun2009,Guo2013}. Phase-field methods are based on models of fluid free energy which goes back to the work of van der Waals \cite{Waals1893}, Gibbs \cite{Gibbs1877} and Cahn $et.$ $al.$ \cite{CahnHilliard1958,CahnAllen1978}. The basic idea for phase-field method is to treat the multiphase fluid as one fluid with variable material properties. An order parameter is employed to characterize the different phases, which varies continuously over thin interfacial layers and is mostly uniform in the bulk phases. Sharp interfaces are then replaced by the thin but non-zero thickness transition regions where the interfacial forces are smoothly distributed. One set governing equations for the whole computational domain can be derived variationally from its energy density field, where the order parameter fields satisfy a advection-diffusion equation (usually the advective Cahn-Hilliard equations) and are coupled to the Navier-Stokes equations through extra reactive stresses that mimic surface tension.\\\\
The classical phase-field model, in the case of two incompressible, viscous Newtonian fluids, is the so-called Model H {\cite{Hohenberg1977}}, which couples fluid flow with Cahn-Hilliard diffusion with a conserved parameter. It has been successfully used to simulate complicated mixing flows involving binary incompressible fluid with the same densities for both components (see \cite{Chella1996} for example). Gurtin $et$ $al.$ \cite{Gurtin1996} re-derived this model in the framework of classical continuum mechanics and showed that it is consistent with the second law of thermodynamics in a mechanical version based on a local dissipation inequality. \\\\
One of the fundamental assumptions when deriving Model H is that, the binary fluid is incompressible, more precisely, its total density as well as the densities for each component are constant. Therefore this model is restricted to the density matched case and cannot be used for the case if the two incompressible fluids have different densities. To treat the problems with small density ratios, a Boussinesq approximation is often used, where the small density difference is neglected except that in the gravitational force. The achieved model maintains thermodynamic consistency (see \cite{Hua2011} as an example). This approach however is no longer valid for large density ratios. Several generalizations of Model H for the case of different densities have been presented and discussed by Lowengrub and Truskinovsky \cite{Lowengrub1998}, Boyer \cite{Boyer2002}, Ding $et$ $al.$ \cite{Ding2007}, Shen and Yang \cite{shen2010}, and most recently by Abels $et$ $al.$ \cite{Abels2012}. Benchmark computations for three of them, namely the models of Boyer \cite{Boyer2002}, Ding $et$ $al.$ \cite{Ding2007}, and Abels $et$ $al.$ \cite{Abels2012}, were carried out by Aland and Voigt \cite{Aland2011}. Antanovskii \cite{Antanovskii1995} derived a quasi-incompressible phase-field model for two-phase flow with different densities. The extended model was presented by Lowengrub and Truskinovsky \cite{Lowengrub1998}, where they employed the pressure rather than density as an independent variable and worked through Gibbs free energy. In their model, the two fluids of different densities are assumed be mixed and compressible along the interfacial region (introducing the quasi-incompressibility into the model). The flow in the interfacial region is in general nonsolenoidal ($\nabla \cdot \bold{v} \ne 0$), resulting in an expansion or contraction flow. Thermodynamic consistency is maintained within the resulting system (quasi-incompressible NSCH) where the Navier-Stokes equations are coupled with the Cahn-Hilliard equations, and the kinetic fluid pressure and variable density were introduced into the chemical potential. Very recently, a numerical method for the quasi-incompressible NSCH system with a discrete thermodynamic law (energy law) was presented by Guo $et.$ $al.$ \cite{Guo--new2014}, where the quasi-incompressibility ( the non-solenoidal velocity) near interfaces was captured. Namely, the numerical results reveal that away from interfaces the fluid is incompressible, while near interfaces waves of expansion and contraction are observed. Very recently, another model of quasi-incompressible fluids for the phase transition simulation was developed \cite{Giesselmann2013}, where a discontinuous Galerkin finite element method is used and studied in \cite{Aki2014}. The model considered differs from the quasi-incompressible NSCH system developed in \cite{Lowengrub1998} in that the volume fraction, rather than the mass concentration, is used as the phase variable. In addition, the two models are derived with different energy functional.\\\\
Another assumption for Model H is that the fluid flow is isothermal. However, for the case that considers thermocapillary (Marangoni) effects, the surface tension gradient is produced by the inhomogeneous distribution of the temperature, so that the system can not be assumed to be isothermal and the transport of temperature field can not be ignored. The extension of Model H in non-isothermal case was presented by Jasnow and Vinals \cite{Jasnow1996}, where, to study the thermocapillary motion of droplets, a constant, externally imposed temperature gradient is considered. Several other works, as mentioned above, have also been devoted to use phase-field method to simulate the dynamics of interface with thermocapillary effects {\cite{Borcia2003,Borcia2004,Sun2009,Guo2013}}. For most of these models, the system equations of flow field (the Navier-Stokes equations with extra stress) and phase-field (the advective Cahn-Hilliard equations) are usually derived from the free energy functional that depend on temperature. The energy equations, however, were not derived together with the system equations. Instead, the classical energy transport equations are incorporated into the system directly, or the temperature fields are assumed to be fixed and the energy equations are not needed. In these treatments, thermodynamic consistency can be hardly achieved. It turns out that the concept of thermodynamic consistency plays an important role for the phase-field modelling. As the phase-field model can be derived through variational procedures, thermodynamic consistency of the model equations can serve as a justification for the model. In addition, it ensures the model to be compatible with the laws of thermodynamics, and to have a strict relaxational behaviour of the free energy, hence the models are more than a phenomenological description of an interfacial problem. In \cite{Antanovskii1995}, Antanovskii presented a phase-field model to study the thermocapillary flow in a gap, where to obtain a free energy that depends on the temperature, the Cahn-Hilliard gradient term associated with the phase-field is introduced into the entropy functional of the system, which leads to a corresponding extra term appearing in the energy equation. The resulting system of equations were derived together through the local balance laws and thermodynamic relations, which maintains thermodynamic consistency. A similar gradient entropy term was considered by Anderson and McFadden \cite{Anderson1996} to study a single compressible fluid with different phases near its critical point. In their work, the phase-field model was derived through a thermodynamic formalism \cite{Sekerka1993} based on entropy generation. Through a similar thermodynamic framework, Verschueren $et$ $al.$ \cite{Verschueren2001} presented a phase-field model for two-phase flow with thermocapillary effects in a Hele-Shaw cell. The system equations maintains thermodynamic consistency, in which the energy equation contains an extra term associated with the variations of the phase-field.\\\\
In present paper, we develop a thermodynamically consistent phase-field model for two-phase flows with thermocapillary effects, which allows the binary incompressible fluid (quasi-incompressible fluid) to have different densities, viscoucities and thermal conductivities for each component. By employing thermodynamic framework used by Anderson and McFadden \cite{Anderson1996}, we first derive a phase-field model for binary compressible flows with thermocapillary effects, where the mass concentration is chosen as the phase variable to label the phases, and the Helmholtz free energy is chosen as the fluid free energy. We then derive the model for binary incompressible flows with thermocapillary effects. Following the work of Lowengrub and Truskinovsky \cite{Lowengrub1998}, we employ the pressure rather than density as the independent variable and thus work with the Gibbs free energy. The equations of both models, including the Navier-Stokes equations with extra stress, an advective Cahn-Hilliard equation and energy equation are derived under a thermodynamic framework. To the best of our knowledge, such a thermodynamically consistent phase-field model for binary incompressible fluid with thermocapillary effects, which allows for different physical properties of each component is new. To validate our model, we first show that thermodynamic consistency are maintained in both models, where the first and second laws of thermodynamics are derived from the model equations. We then analyze the model in the sharp-interface limit to show that the governing equations and interfacial conditions of the classical sharp-interface model can be recovered from our phase-field models, which reveals the underlying physical mechanisms of phase-ﬁeld model. In the jump condition of the momentum balance, we relate the surface tension term of our phase-field model to that of the classical sharp-interface model by introducing a ratio parameter, where the value of the parameter can be determined through the relation. As another validation of our model, two examples are computed by using a continuous finite element method, including thermcapillary convection in two-layer fluid system and thermocapillary migration of a bubble in a medium fluid. The numerical results for the first two examples are consistent with the corresponding analytical solutions \cite{Pendse2010} and the existing numerical solutions \cite{Herrmann2008}. Note that for all the examples computed in this paper, we assume that the interface have no contact with the boundary of the domain. In the case that the interface contacts with the boundary of the domain, extra difficulties would arise from complicated interface/boundary interacting conditions and should be dealt with separately (e.g., \cite{Qian2006, Eck2009,Gao2012}).\\\\
The paper is organized as follows. In $\S$2, we introduce the variable density and mass-averaged velocity of the binary fluid. We then present the derivations of the phase-field model for binary compressible fluid with thermocapillary effects in $\S$3, and the corresponding derivations for the binary incompressible (quasi-incompressible) fluid in $\S$4. The sharp-interface limit analysis of our phase-field model is carried out in $\S$5. $\S$6 shows some numerical results as validations of our model. Finally, conclusion and future work are discussed in $\S$7.
\Section{Variable density and mass-averaged velocity}
In phase-field modelling, an order parameter (phase variable) is normally introduced to distinguish different phases and the intervening interface. Lowengrub \& Truskinovsky \cite{Lowengrub1998} have argued for the advantage of using a physically realistic scalar field instead of an artificial smoothing function for the interface. Several physically realistic scalar fields have been suggested as the order parameters for phase-field modelling, e.g. the mass density $\rho$ for the case of a single compressible fluid with different phases \cite{Anderson1996}, the mass concentration $c$ of one of the constituents for the case of compressible and incompressible binary fluid \cite{Lowengrub1998, Abels2012}, or an alternative phase variable, the volume fraction $\phi$ for the case of incompressible binary fluid \cite{LiuShen2002} and solidification of single materials \cite{Wang1993}. Here we choose the mass concentration $c$ of one of the constituents as the phase variable, and begin by introducing the variable density for the mixture. We consider a mixture of two fluids in a domain $\Omega$, and take a sufficient small material volume $V \in \Omega$. We then have the following theorem (e.g. \cite{continuum}),
\begin{theo}\label{theo1}
For a smooth function $f(\bold{x},t)$  in the Eulerian coordinate, 
\begin{equation}
\frac{{\rm d}}{{\rm d}t}\int_{V(t)}f(\bold{x},t){\rm d}V=\int_{V(t)}\bigg(\frac{{\rm D}f}{{\rm D}t}+f(\nabla \cdot \bold{v})\bigg){\rm d}V=\int_{V(t)}\bigg(\frac{\partial f}{\partial t}+\nabla \cdot\big(f \bold{v}\big)\bigg){\rm d}V,
\end{equation}
where ${\rm D}/{\rm D}t = \partial /\partial t + \bold{v}\cdot\nabla $ is the material derivative and $\bold{v}$ is the velocity of the moving volume $V(t)$.
\end{theo}
In the control volume, the two fluids are labeled by $i=1,2$ and they fill the volumes $V_i$ separately. We then introduce the volume fraction $\gamma_i$ for the $i$th fluid such that $\gamma_i={V_i}/{V}$. Further we assume that two fluids can mix along the interfacial region and the volume occupied by a given amount of mass of the single fluid does not change after mixing. Then within the material volume $V$, $\gamma_i$ satisfy the condition $\gamma_1+\gamma_2=1$. Let $M=M_1+M_2$ be the total mass of the mixture, and $M_i$ be the mass of the $i$th fluid in the volume. We now introduce the local volume-averaged mass density taken over the sufficient small volume $V$ for each fluid $\tilde{\rho}_i={ M_i}/{V}$, and the actual local mass density for each fluid $\rho_i = {M_i}/{V_i}$. Note that for incompressible components, we assume that $\rho_i$ are uniform constants. Having in mind the definition of volume fraction, we obtain the relation between the volume-averaged mass densities and the local mass densities
\begin{equation}
\gamma_i=\frac{\tilde{\rho_i}}{\rho_i}~~~~{\rm and }~~~~\frac{\tilde{\rho_1}}{\rho_1}+\frac{\tilde{\rho_2}}{\rho_2}=1.\label{volume--frac2}
\end{equation}
We then define the volume-averaged mass density for the mixture as
\begin{equation}
\rho=\tilde{\rho}_1+\tilde{\rho}_2=\frac{M_1+M_2}{V}=\frac{M}{V}.\label{volume--frac5}
\end{equation}
Let $c_i$ be the mass concentration for the $i$th fluid, such that
\begin{equation}
c_i=\frac{M_i}{M}=\frac{\tilde{\rho}_i}{\rho}~~~~{\rm and}~~~~c_1+c_2=1.\label{volume--frac6}
\end{equation}
Using Eqs.(\ref{volume--frac2}) and (\ref{volume--frac6}), we obtain the variable density for the mixture of two fluids
\begin{equation}
\frac{1}{\rho(c)}=\frac{c}{\rho_1}+\frac{1-c}{\rho_2}.\label{variable--c}
\end{equation}
Here we chose the mass concentration of fluid 1 as the phase variable for our phase-field model, such that $c=c_1=1-c_2$. It can be seen that, for two incompressible components of different densities, the variable density $\rho(c)$ for the mixture is constant almost everywhere except near the interfacial region. For simplicity, we write the variable density $\rho(c)$ as $\rho$ in all the following derivations.\\
Now we suppose that the two fluids move with different velocities $\bold{v}_i(\bold{x},t)$. The equation of mass balance for each fluid within the material volume $V$ can then be written in the form \cite{Lowengrub1998,Boyer2002,Abels2012}
\begin{equation}
\frac{\partial \tilde{\rho}_i }{\partial t}+\nabla \cdot(\tilde{\rho}_i  \bold{v}_i)=0.\label{mass-fluid1--v1}
\end{equation}
We then introduce the mass-averaged velocity for the mixture as
\begin{equation}
\rho\bold{v}=\tilde{\rho}_1\bold{v}_1+\tilde{\rho}_2\bold{v}_2~~~~{\rm or}~~~~\bold{v}=c_1\bold{v}_1+c_2\bold{v}_2.\label{mass-averaged-vel}
\end{equation}
Substituting the density (\ref{volume--frac5}) and mass-averaged velocity (\ref{mass-averaged-vel}) into Eq. (\ref{mass-fluid1--v1}), we obtain the mass balance for the mixture of two fluids
\begin{equation}
\frac{\partial \rho }{\partial t}+\nabla \cdot(\rho  \bold{v})=0.\label{mass-cons-ab}
\end{equation}
In the following derivations, we consider the mixture as a single fluid moving with velocity $\bold{v}$. Note that if we consider a binary incompressible fluid (assuming the two fluids of the mixture are incompressible, and the temperature effects on the densities of both fluids are negligible), then $\rho_1$ and $\rho_2$ are constants, and the above equation (\ref{mass-cons-ab}) can be further written as 
\begin{equation}
\nabla \cdot \bold{v}=-\frac{1}{\rho}\frac{{\rm D} \rho}{{\rm D}t}  =-\frac{1}{\rho}\frac{{\rm d} \rho}{{\rm d} c} \frac{{\rm D} c}{{\rm D}t}=\alpha\rho\frac{{\rm D} c}{{\rm D}t},\label{quasi-incompressibility}
\end{equation}
where $\alpha=(\rho_2-\rho_1)/\rho_2\rho_1$ is constant. We note that, due to the variations of the phase variable $c$, the mass-averaged velocity for the mixture is non-solenoidal ($\nabla \cdot \bold{v}\ne 0$) near the interfacial region, which introduces the compressibility effects into the model. Such binary incompressible fluid is termed as the quasi-incompressible fluid (e.g. \cite{Antanovskii1995, Lowengrub1998}).\\
We remark that except for this mass-averaged velocity $\bold{v}$, another velocity for the mixture, the volume-averaged velocity $\tilde{\bold{v}}$ was considered in \cite{Abels2012,Boyer2002}, and \cite{Ding2007}, where the volume fraction $\gamma$ instead of the mass concentration $c$ is employed as the phase variable, and further used to relate the velocity of single fluids and mixture. This volume-averaged velocity of binary incompressible fluid is solenoidal ($\nabla \cdot \tilde{\bold{v}}=0$) over the whole domain, where an extra term that accounts for the mass flux relative to the volume-averaged velocity appears in the Navier-Stokes equations (see, for details, \cite{Abels2012}).
\Section{Phase-field model for binary compressible fluid with thermocapillary effects}
In this section, we develop a system of equations for a binary fluid with thermocapillary effects, in which both components are compressible and Cahn-Hilliard diffusion is coupled with fluid motion.
\subsection{Derivation of the model}
We first consider a mixture of two fluids in a domain $\Omega$, and we take an arbitrary material volume $V \in \Omega$ that moves with the mixture. Within the material volume, we define the properties for the binary compressible fluid as
\begin{align}
M&=\int_{V(t)} \rho  ~{\rm d}V,\label{vol--com--mass}\\
\bold{P}&=\int_{V(t)}\rho \bold{v} ~{\rm d}V,\label{vol--com--mom}\\
E&=\int_{V(t)}\bigg(\frac{1}{2} \rho | \bold{v} |^2+ \rho g  z +\rho \hat{u} \bigg){\rm d}V,\label{vol--com--en}\\
S&=\int_{V(t)}\rho \hat{s}{~\rm d}V,\label{vol--com--entro}\\
C&=\int_{V(t)}\rho   c  ~{\rm d}V,\label{vol--com--phi}
\end{align}
where $M$, $\bold{P}$, $E$, $S$ is the total mass, momentum, energy, and entropy of the mixture, $\rho(c)$ is the variable density of the mixture, $\bold{v}$ is the mass-averaged velocity of the mixture, $|\bold{v}|^2/2$ is the kinetic energy per unit mass, $g z$ is the gravitational potential energy per unit mass, $z$ is the z-coordinate, $\hat{u}$ ($\hat{s}$) is the internal energy (entropy) per unit mass, $c$ is the phase variable. Substituting the mass concentration (\ref{volume--frac6}) into Eq.(\ref{vol--com--phi}) gives
\begin{equation}
C=\int_{V(t)}\rho c~{\rm d}V=\int_{V(t)}\rho c_1~{\rm d}V=\int_{V(t)}\tilde{\rho}_1~{\rm d}V,\label{mass--fluid1}
\end{equation}
where $C$ stands for the constituent mass of fluid $1$ within the material volume $V(t)$. In phase-field modelling, except the classical free energy density for bulk phases, an extra gradient term is typically added into the model accounting for the free energy of the diffuse interface \cite{CahnHilliard1958}. Several ways have been suggested to introduce the gradient term into the phase-field model, e.g. by introducing it into the entropy functional \cite{Antanovskii1995,Wang1993}, free energy functional \cite{Lowengrub1998} or internal energy functional \cite{Anderson1998, Verschueren2001}. In the present work, as the thermocapillary effects along the interface are investigated, we expect that the surface free energy (serving as the surface tension (See $\S 5.4$)) of our phase-field model is temperature dependent. Therefore, according to the thermodynamic relations, we introduce the gradient term into both the internal energy and entropy of our model, such that
\begin{eqnarray}
&&\hat{u}(s,\rho, c, \nabla c)=u(s,\rho,  c)+u^{grad}(\nabla   c),~~~~~~~~~u^{grad}=\lambda _{u}\frac{1}{2}|\nabla   c|^2, \label{def--internal--com}\\
&&\hat{s}(T,\rho, c, \nabla c)=s(T,\rho,  c)+s^{grad}(\nabla   c),~~~~~~~~~s^{grad}=\lambda_{s}\frac{1}{2}|\nabla   c|^2,\label{internal--energy}\\
&&\hat{f}(T,\rho, c, \nabla c)=f(T,\rho,  c)+f^{grad}(T,\nabla   c),~~~~f^{grad}=\lambda_{f}(T)\frac{1}{2}|\nabla   c|^2,\label{internal--free--energy}
\end{eqnarray}
where $u$, $s$ and $f$ stand for the classical parts of the specific internal energy, entropy and free energy separately. Here $f$ is the Helmholtz free energy. The parts $u^{grad}$, $s^{grad}$ and $f^{grad}$ are the gradient terms analogous to the Landau-Ginzburg \cite{GinzburgLandau1950} or Cahn-Hilliard \cite{CahnHilliard1958} gradient energy. Note that these parts are termed as the "non-classical" terms by Anderson $et$ $al.$ \cite{Anderson1996} who used a phase-field model to study a single compressible fluid with different phases near its critical point. In addition, $\lambda_{u}$ and $\lambda_{s}$ are constant parameters, $\lambda_f(T)$ is a parameter depending on the temperature and will lead to the thermocapillary effects along the interface. Note that $\lambda_{u}$, $\lambda_{s}$ and $\lambda_{f}(T)$ can be further used to relate the surface tension of the phase-field model to that of the sharp-interface model when the phase-field model reduces to its sharp-interface limit (see $\S 5.4$ for details). As $u(\rho,s, c)$ is the classical contribution to the specific internal energy $\hat{u}$, we have the thermodynamic relation
\begin{align}
{\rm d} u (s,\rho,  c) = \frac{\partial u}{\partial s} \bigg|_{\rho,  c}{\rm d}s +\frac{\partial u}{\partial \rho}\bigg|_{s,  c}{\rm d} \rho +  \frac{\partial u}{\partial   c} \bigg|_{s,\rho}{\rm d}  c =  T {\rm d}s +\frac{p}{\rho^2}{\rm d}\rho +  \frac{\partial u}{\partial   c}\bigg|_{s,\rho} {\rm d}  c,\label{com-der-e}
\end{align}
where the subscripts indicate which variables are held constant when the various partial derivatives are taken. This relation states that the heat transfer ($T{\rm d}s$), pressure-volume work $(p/\rho^2{\rm d}\rho$) and chemical work ($({\partial u}/{\partial c} ){\rm d}c$) all contribute to the changes in the internal energy. Further, we have the thermodynamic relation for Helmholtz free energy
\begin{equation}
f=u-Ts.\label{rela--free1}
\end{equation}
Having in mind the relation (\ref{com-der-e}), we obtain
\begin{equation}
{\rm d}f={\rm d}u-{\rm d}(Ts)={\rm d}u-s{\rm d}T-T{\rm d}s= \frac{p}{\rho^2}{\rm d}\rho -s{\rm d}T+  \frac{\partial u}{\partial   c} \bigg|_{s,\rho}{\rm d}  c,\label{rela--free2}
\end{equation}
such that
\begin{equation}
\frac{\partial f}{\partial \rho}\bigg|_{T,  c}=\frac{p}{\rho^2},~~~~\frac{\partial f}{\partial T}\bigg|_{\rho,  c}=-s~~~~{\rm and}~~~~\frac{\partial f}{\partial c}\bigg|_{T, \rho}=\frac{\partial u}{\partial c}\bigg|_{s,\rho}.\label{rela--free3}
\end{equation}
Similarly, we assume that the same thermodynamic relations, which hold for the classical terms also hold for the general terms, such that
\begin{equation}
\hat{f}=\hat{u}-T\hat{s}~~~~{\rm and}~~~~\frac{\partial \hat{f}}{\partial T}\bigg|_{s, \rho,   c, \nabla   c}=-\hat{s}.\label{helm--free--sub}
\end{equation}
With the relations (\ref{rela--free1}) and (\ref{rela--free3}), we must also have the relations for the gradient terms
\begin{equation}
f^{grad}=u^{grad}-Ts^{grad}~~~~{\rm and}~~~~\frac{\partial f^{grad}}{\partial T}\bigg|_{\nabla   c}=-s^{grad},
\end{equation}
and for the corresponding coefficients
\begin{equation}
\lambda_f(T) = \lambda_u-T\lambda_s~~~~{\rm and}~~~~\frac{{\rm d} \lambda_f(T)}{{\rm d} T}=-\lambda_s.\label{con--free--1}
\end{equation}
For simplicity, we omit all the subscripts in the following derivations. Under the assumptions above, the general forms of physical balance associated with $M$, $\bold{P}$, $E$, $S$ and $C$ can be given as follows
\begin{align}
\frac{{\rm d} M}{{\rm d} t}&=0,\label{der--com--mass}\\
\frac{{\rm d} \bold{P}}{{\rm d} t}&=\int_{\partial V (t)} \bold{m} \cdot \hat{\bold{n}} ~{\rm d}A - \int_{V(t)}\rho g \hat{\bold{z}}~ {\rm d}V,\label{der--com--mom}\\
\frac{{\rm d} E}{{\rm d} t}&=\int_{\partial V (t)} \bigg(\bold{v} \cdot \bold{m} \cdot \hat{\bold{n}} - \bold{q}_E\cdot \hat{\bold{n}}  - \bold{q}_{E}^{nc} \cdot \hat{\bold{n}} \bigg) {\rm d}A,\label{der--com--en}\\
\frac{{\rm d} S}{{\rm d} t}&=- \int_{\partial V (t)} \bigg( \frac{\bold{q}_E}{T} \cdot \hat{\bold{n}} + \bold{q}_{S}^{nc}\cdot \hat{\bold{n}}\bigg){\rm d}A+\int_{V(t)} S_{gen}~{\rm d}V ~~~~\big(S_{gen}\geqslant 0\big),\label{der--com--entro}\\
\frac{{\rm d}  C}{{\rm d} t}&= -\int_{\partial V (t)} \bold{q}_{C}\cdot \hat{\bold{n}}~{\rm d}A,\label{der--com--phi}
\end{align}
where $\hat{\bold{n}}$ is the unit outward normal vector of the boundary, $\hat{\bold{z}}$ is the vertical component of the unit normal vector. Eq.(\ref{der--com--mass}) represents the mass balance of the mixture within the volume. Eq.(\ref{der--com--mom}) represents the momentum balance, stating that the rate of the change in total momentum equals to the force (surface forces $\bold{m}$ and body forces $\rho g \hat{\bold{z}}$) acting on the volume. Here only the gravitational forces are considered. The energy balance equation (\ref{der--com--en}) states that the change in total energy equals to the rate of work done by the forces ($\bold{m}$) on the boundary plus the energy flux (classical $\bold{q}_{E}$ and non-classical $\bold{q}^{nc}_{E}$ internal energy flux) through the boundary. The entropy balance (\ref{der--com--entro}) states that the rate of change of entropy in the control volume during the process equals to the net entropy transfer through the boundary (classical $\bold{q}_{E}/T$ and non-classical $\bold{q}^{nc}_{S}$ entropy flux) plus the local entropy generation ($S_{gen} \geqslant0$) within the control volume (e.g. \cite{Moran2010}). Based on the second law of thermodynamics, the local entropy generation is non-negative for a dissipative system (or say for an irreversible process), which is key to the thermodynamic frame that we used for the derivations. For the constituent mass balance (\ref{der--com--phi}), we use Eq.(\ref{mass--fluid1}) and Theorem \ref{theo1} to obtain
\begin{eqnarray}
\frac{{\rm d}C}{{\rm d}t}=\frac{\rm d}{{\rm d} t}\int_{V(t)}\tilde{\rho}_1{\rm d}V=\int_{V(t)}\bigg(\frac{\partial \tilde{\rho}_1}{\partial  t}+\nabla \cdot (\tilde{\rho}_1\bold{v})\bigg){\rm d}V=-\int_{\partial V (t)} \bold{q}_{C}\cdot \hat{\bold{n}}~{\rm d}A.\label{mass--fluid11}
\end{eqnarray}
Substituting (\ref{mass-fluid1--v1}) into (\ref{mass--fluid11}), we obtain $\bold{q}_{C}=\tilde{\rho}_1(\bold{v}_1- \bold{v})$, where $\bold{q}_{C}$ stands for the mass flux of fluid 1 with velocity $(\bold{v}_1-\bold{v})$ through the boundary of control volume. Note that in the following derivations $q_C$ will be related to the chemical potential of the phase field, which is analogous to the standard derivations for the Cahn-Hilliard equations (see, for examples, \cite{Anderson1998} and \cite{Lowengrub1998}).\\
In what follows, we use the definitions (\ref{vol--com--mass})-(\ref{vol--com--phi}) and the balance laws (\ref{der--com--mass})-(\ref{der--com--phi}) to obtain the equations that expressed in terms of the above unknowns, including $\bold{m}$, $\bold{q}_{E}$, $\bold{q}^{nc}_{E}$, $\bold{q}^{nc}_{S}$, $\bold{q}_{C}$ and $S_{gen}$. We then specify those unknowns with respect to the second law of thermodynamics (ensuring $S_{gen}\geqslant 0$) and the concept of thermodynamic consistency of the phase-field model.\\
For mass balance (\ref{der--com--mass}), we use Theorem \ref{theo1} to obtain
\begin{eqnarray}
\frac{{\rm D} \rho}{{\rm D} t} = - \rho (\nabla \cdot \bold{v}),\label{equ--com--mass}
\end{eqnarray}
based on which, we have the following,
\begin{theo}\label{theo2}
{\rm (Transport Theorem 2)}
For a smooth function $f(\bold{x},t)$  in the Eulerian coordinate, 
\begin{eqnarray}
\frac{{\rm d}}{{\rm d}t}\int_{V(t)}\rho f(\bold{x},t)~{\rm d}V=\int_{V(t)}\rho\frac{{\rm D}f}{{\rm D}t}~{\rm d}V=\int_{V(t)}\rho\bigg(\frac{\partial f}{\partial t}+(\bold{v}\cdot \nabla ) f \bigg){\rm d}V,
\end{eqnarray}
where $\rho$ is the density of the mixture defined in the volume $V(t)$ and satisfies the mass balance (\ref{equ--com--mass}).
\end{theo}
Note that as Theorem \ref{theo1} and Theorem \ref{theo2} are frequently used, we will not refer them in the following derivations.\\
For momentum balance (\ref{der--com--mom}), we simply have
\begin{eqnarray}
\rho \frac{{\rm D} \bold{v}}{{\rm D} t}=\nabla \cdot \bold{m}-\rho g \hat{\bold{z}}.\label{equ--com--mom}
\end{eqnarray}
For energy balance (\ref{der--com--en}), we obtain
\begin{align}
\rho T \frac{{\rm D} s}{{\rm D} t} &=-\nabla \cdot (\rho \lambda_{u} \frac{{\rm D}   c}{{\rm D} t}\nabla   c) +\big(\bold{m} -\rho\lambda_{u} |\nabla c|^2 \bold{I}+\rho \lambda_{u}(\nabla   c \otimes \nabla   c)\big) : \nabla \bold{v}\nonumber\\
&+ \lambda_{u}\rho  \Delta   c \frac{{\rm D}    c}{{\rm D} t}  -\nabla \cdot  \bold{q}_E - \nabla \cdot \bold{q}^{nc}_E- \rho \frac{\partial u}{\partial   c} \frac{{\rm D}   c}{{\rm D} t} -\frac{p}{\rho}\frac{{\rm D}\rho}{{\rm D}t}.\label{equ--com--en--4}
\end{align}
where Eqs.(\ref{com-der-e}), (\ref{equ--com--mass}) and (\ref{equ--com--mom}), and the following identities are used
\begin{eqnarray}
\frac{{\rm d}}{{\rm d}t}\int_{V(t)}\rho g z~{\rm d}V=\int_{V(t)}\rho g \bold{v}\cdot \nabla z~{\rm d}V=\int_{V(t)}\rho g \bold{v}\cdot \bold{\hat{z}}~{\rm d}V,
\end{eqnarray}
and
\begin{eqnarray}
\rho\frac{{\rm D}}{{\rm D} t}(\frac{1}{2}\lambda_{u}|\nabla  c |^2)=\nabla \cdot (\rho \lambda_{u} \frac{{\rm D}   c}{{\rm D} t}\nabla   c) +\big(\rho\lambda_{u} |\nabla c|^2 \bold{I}-\rho  \lambda_{u} (\nabla   c \otimes \nabla   c)\big): \nabla \bold{v}  - \rho\lambda_{u} \Delta c \frac{{\rm D}    c}{{\rm D} t}.
\label{id--com--en--1}
\end{eqnarray}
Here $``:"$ stands for the double dot product of the stress tensor (e.g. \cite{continuum}).\\
For entropy balance (\ref{der--com--entro}), we obtain
\begin{align}
 \rho\frac{{\rm D} s}{{\rm D} t}&=- \nabla \cdot (\rho \lambda_{s} \frac{{\rm D} c}{{\rm D} t}\nabla   c) +\big(-\rho\lambda_{s} |\nabla c|^2 \bold{I} +\rho  \lambda_{s}(\nabla c \otimes \nabla c)\big): \nabla \bold{v}\nonumber\\
& +\lambda_{s}\rho\Delta   c \frac{{\rm D}    c}{{\rm D} t}-\nabla \cdot (\frac{\bold{q}_{E}}{T})+S_{gen}-\nabla \cdot \bold{q}_S^{nc},\label{equ--com--entro--1}
\end{align}
where, similar to Eq.(\ref{id--com--en--1}), the following identity is used,
\begin{align}
\rho\frac{{\rm D}}{{\rm D} t}(\frac{1}{2}\lambda_{s}|\nabla  c |^2)=\nabla \cdot (\rho \lambda_{s} \frac{{\rm D}   c}{{\rm D} t}\nabla   c)+\big(\rho\lambda_{s} |\nabla c|^2 \bold{I}-\rho  \lambda_{s} (\nabla   c \otimes \nabla   c)\big) : \nabla \bold{v}- \rho\lambda_{s}\Delta   c \frac{{\rm D}    c}{{\rm D} t}.\nonumber\\
\end{align}
For constituent mass balance (\ref{der--com--phi}), we simply have
\begin{eqnarray}
\rho \frac{{\rm D}   c}{{\rm D} t}= - \nabla \cdot \bold{q}_{C}.\label{equ--com--phi}
\end{eqnarray}
We then use Eq.(\ref{equ--com--entro--1}) and Eq.(\ref{equ--com--phi}) to substitute the terms $\rho{\rm D}s/{\rm D}t$ and $\rho{\rm D}c/{\rm D}t$ in (\ref{equ--com--en--4}), and use the relation (\ref{con--free--1}) to obtain the expression for the entropy generation,
\begin{align}
S_{gen}=&\frac{1}{T}\bigg(\bold{m}-\rho  \lambda_{f}( T)|\nabla c|^2\bold{I}+\rho \lambda_{f} (T)\bold{T} + p\bold{I} \bigg) : \nabla \bold{v}+\nabla \frac{1}{T}  \cdot \bigg(  \rho \lambda_{u}\frac{{\rm D}   c}{{\rm D} t}\nabla   c+ \bold{q}_E + \bold{q}^{nc}_E\nonumber\\
&-\mu_{C} \bold{q}_{C}\bigg)-\nabla \cdot \bigg[\frac{1}{T}\rho \lambda_{f}(T)  \frac{{\rm D}   c}{{\rm D} t}\nabla   c +\frac{1}{T}\bold{q}^{nc}_E-\frac{1}{T}\mu_{C} \bold{q}_{C}-\bold{q}_S^{nc}\bigg]-\frac{1}{T}\bold{q}_{C}\cdot\nabla \mu.\label{com--entropy}
\end{align}
To ensure the non-negativity of the entropy generation $S_{gen}\geqslant 0$ (second law of thermodynamics), we specify the unknown terms in the form
\begin{eqnarray}
&&\bold{q}_E=-k(c)\nabla T,\hspace{26mm}\bold{q}_{E}^{nc}=-\rho \lambda_{u}\frac{{\rm D}  c}{{\rm D} t} \nabla c+\mu_{C} \bold{q}_{C},\label{relation-qe}\\
&&\bold{q}_{S}^{nc}=-\rho \lambda_{s}\frac{{\rm D}   c}{ {\rm D} t} \nabla   c,\hspace{10mm}\bold{m} =\rho  \lambda_{f}( T)|\nabla c|^2\bold{I}-\rho\lambda_{f}(T)\bold{T} +\boldsymbol{\sigma},\\
&&\mu_{C} =\frac{\partial f}{\partial c}-\lambda_{f}(T)\Delta c, \hspace{34mm}\bold{q}_{C}=-m_C\nabla \mu_{C},\label{relation-mu}\\
&&\bold{T} =\nabla c\otimes \nabla c,\hspace{50.5mm} \boldsymbol{\sigma}=-p\bold{I}+\boldsymbol{\tau},\\
&& p=\rho^2\frac{\partial f}{\partial \rho}, \hspace{18.5mm}\boldsymbol{\tau}=\mu(c)(\nabla \bold{v} +\nabla \bold{v}^{T} )-\frac{2}{3}\mu(c) (\nabla \cdot \bold{v})\bold{I}.\label{relation-tau}
\end{eqnarray}
Note that $\boldsymbol{\tau}$ is the deviatoric stress tensor from the classical Navier-Stokes equations (e.g. \cite{Batchelor2000}). Here we use the thermodynamic relation (\ref{rela--free3}) to obtain the chemical potential $\mu_{C}$. The pressure $p$ can be obtained immediately through the thermodynamic relation (\ref{rela--free3}). By substituting the above terms into (\ref{der--com--mass})-(\ref{der--com--phi}), we obtain the system of equations for the phase-field model governing binary compressible flows with thermocapillary effects
\begin{align}
\frac{{\rm D} \rho}{{\rm D} t} &= - \rho (\nabla \cdot \bold{v}),\label{sys--com--mass}\\
\rho \frac{{\rm D} \bold{v}}{{\rm D} t}&=\nabla \cdot \bold{m}-\rho g \hat{\bold{z}},\label{sys--com--mom}\\
\rho \frac{{\rm D} u}{{\rm D} t}&=(\boldsymbol{\sigma}-\rho T\lambda_{s}|\nabla c|^2\bold{I}+\rho T\lambda_{s}\bold{T}) : \nabla \bold{v}  +\lambda_{u}\nabla \cdot (\rho  \nabla   c) \frac{{\rm D}    c}{{\rm D} t}+\nabla \cdot  (k(c)\nabla T\nonumber\\
&+m_{C}\mu_{C}\nabla \mu _{C}),\label{sys--com--en}\\
\rho \frac{{\rm D} s}{{\rm D} t}&=\frac{1}{T}(\boldsymbol{\tau} -\rho T\lambda_{s}|\nabla c|^2\bold{I}+\rho T\lambda_{s}\bold{T}) : \nabla \bold{v}+ \lambda_{s}\nabla \cdot (\rho  \nabla   c) \frac{{\rm D}    c}{{\rm D} t}+ \frac{1}{T}\nabla \cdot (k(c)\nabla T),\nonumber\\
& \label{sys--com--entro}\\
\rho \frac{{\rm D}   c}{{\rm D} t}&=m_C\Delta \mu_{C},\label{sys--com--phi1}\\
\mu_{C} &=\frac{\partial f}{\partial c}-\lambda_{f}(T) \Delta   c.\label{sys--com--phi2}
\end{align}
Note that the second term in the stress tensor $\bold{m}$ is the extra reactive stress (Ericksen's stress) to mimic the surface tension. This stress term is associated with the presence of concentration gradients energy (Cahn-Hilliard energy). We note that the temperature dependent coefficient in $\bold{m}$ is a linear function of temperature, which leads to the thermocapillary effects along the interface (see $\S 4.2$ for details). $m_C$ is a positive constant standing for the mobility of the diffuse interface. Note that in the non-classical heat (or entropy) flux $\bold{q}_E^{nc}$ (or $\bold{q}_S^{nc}$), the term $\rho \lambda_u \nabla c~{\rm D}c/{\rm D }t$ (or $\rho \lambda_s \nabla c~{\rm D}c/{\rm D }t$) is associated with the gradient energy (or entropy, respectively) and is in the direction of the gradient of the phase variable. Similar terms were obtained by Wang $et$ $al.$ \cite{Wang1993} who used a phase-field model to study the solidification of single material, and by Anderson $et$ $al.$ \cite{Anderson1996} who used a phase-field model to study a single compressible fluid with different phases near its critical point. In addition, a non-classical energy flux term $m_C\mu_{C}\nabla \mu_{C}$ appears in our energy balance equation (\ref{sys--com--en}). The same energy flux term was obtained by Gurtin $et$ $al.$ \cite{Gurtin1996} (see Eq.(28)), who re-derived the Model H in the framework of classical continuum mechanics. A ``counterpart'' entropy flux term was identified by Lowengrub $et$ $al.$ \cite{Lowengrub1998} when deriving a phase-field model for binary compressible fluid, where this term is required to keep the model compatible with the second law of thermodynamics. In the latter work, the isothermal fluid flow was studied, so that the temperature $T$ in the entropy flux was treated as constant, whereas in our work, the temperature is not a constant as the thermocapillary effects are considered here. They identified this non-classical as the entropy flux transported through the boundary by chemical diffusion. Our model agrees with these works well and therefore we identify this non-classical energy flux term as the energy that carried into the control volume by the chemical diffusion. Note that several phase-field models (e.g., \cite{Blesgen1999, Allarire2002, Abels2007}) have been presented to study the binary compressible fluids, where the specifications of free energy (Eq.(\ref{relation-tau})) that contribute to the compressibility of the binary compressible fluids are discussed and provided.
\\
Similar to the approach that defines the variable density $\rho(c)$ (\ref{variable--c}), we define the variable viscosity $\mu(c)$ and the variable thermal diffusivity $k(c)$ for the mixture in the form of the harmonic average,
\begin{eqnarray}
\mu(c)=\frac{\mu_1 \mu_2}{(\mu_2-\mu_1)c+\mu_1},~~~~k(c)=\frac{k_1 k_2}{(k_2-k_1)c+k_1},\label{variable--thermaldiffusicity}
\end{eqnarray}
where $\mu_i$,and $k_i$ are the viscosity and thermal conductivity of the $i$th fluid.
\subsection{Thermodynamic consistency and Galilean invariance} 
As our phase-field model (\ref{sys--com--mass})-(\ref{sys--com--phi2}) is derived within a thermodynamic framework, it implies that the first and second thermodynamic laws are naturally underlying the model. However, from the numerical point of view, thermodynamic consistency can be further served as a criterion to design the numerical methods. In our phase-field model, the Navier-Stokes equations are coupled with the Cahn-Hilliard equations and energy balance equation, which leads to a nonlinear system. Moreover, as the rapid variations in the solutions of the phase variable occur near the interfacial region, the energy stability of the numerical method is critical. Recently, the preservation of the thermodynamic laws at discrete level has been reported to play an important role in the designing of numerical methods (e.g. \cite{Lin2006,Lin2007} for liquid crystal models, \cite{Hua2011, Guo2014} for phase-field models), which not only immediately implies the stability of the numerical scheme, but also ensures the correctness of the solutions. Hence, in contrast to the derivations, we now show that the first and second laws of thermodynamics can be derived from the system of equations (\ref{sys--com--mass})-(\ref{sys--com--phi2}), which can be further used to design the numerical methods. In addition, important modelling properties Onsager reciprocal relations and Galilean invariance will be verified as well.
\subsubsection{The laws of thermodynamic}
Multiplying Eqs. (\ref{sys--com--mass}), (\ref{sys--com--mom}) and (\ref{sys--com--entro})-(\ref{sys--com--phi2}) by $p/\rho+\bold{v}\cdot \bold{v}/2+u$, $\bold{v}$, $T$, $\mu_{C}$ and $\rho {\rm D}  c/{\rm D}t$, and summing them up, we can obtain the first law of thermodynamics (\ref{der--com--en}) that we used to derive the model. By substituting the terms, $\bold{m}$, $\bold{q}_{E}$, $\bold{q}^{nc}_{E}$, $\bold{q}^{nc}_{S}$ and $\bold{q}_{C}$ into the entropy generation (\ref{com--entropy}), we obtain the second law of thermodynamics,
\begin{align}
S_{gen} &=\frac{1}{T}\boldsymbol{\tau} : \nabla \bold{v} - \bold{q}_{E}\frac{\nabla T}{T^2}-\bold{q}_{C}\frac{1}{T}\nabla \mu_{C} \nonumber\\
&=\frac{1}{T}\boldsymbol{\tau} : \nabla \bold{v} + k(c)|\frac{\nabla T}{T}|^2+\frac{m_C}{T}|\nabla \mu_{C}|^2 \geqslant 0,\label{entro--com233}
\end{align}
where we see that the viscous dissipation, heat transfer and chemical potential (the variation of the phase variable) all contribute to the entropy generation of our phase-field model. Note that the same entropy generation equation was obtained by Lowengrub \& Truskinovsky \cite{Lowengrub1998} when deriving the phase-field model for the binary compressible fluid.
\subsubsection{Onsager~reciprocal~relations}
From Eq.(\ref{entro--com233}), we observe that the entropy generation can be seen as the sum of terms each being a product of a flux ($\boldsymbol{\tau}$, $\bold{q}_{E}$, $\bold{q}_C$) and thermodynamic forces ($\nabla \bold{v}$, $\nabla T$, $\nabla \mu_{C}$). The simplest model, based on the linear thermodynamics of non-equilibrium processes \cite{BookGroot1985}, assumes linear relations between the fluxes and thermodynamic forces, such that 
\begin{align}
\boldsymbol{\tau}& = L_{11}\nabla \bold{v}+ L_{12}\nabla T+ L_{13}\nabla \mu_{C},\nonumber\\
\bold{q}_E &= L_{21}\nabla \bold{v}+ L_{22}\nabla T+ L_{23}\nabla \mu_{C},\nonumber\\
\bold{q}_C &= L_{31}\nabla \bold{v}+ L_{32}\nabla T +L_{33}\nabla \mu_{C},
\label{relation--linear}
\end{align}
where the coefficients $L_{ij}$ are chosen to guarantee the non-negativity of $S_{gen}$. Moreover,  microscopic reversibility requires the Onsager reciprocal relations $L_{ij}=L_{ji}$ (\cite{BookGroot1985}, and see, for examples, \cite{Qian2006, Eck2009}). From Eqs.(\ref{relation-tau}), (\ref{relation-qe}) and (\ref{relation-mu}), we see that our choices of $\boldsymbol{\tau}, \bold{q}_E$ and $\bold{q}_C$ satisfy the linear relation (\ref{relation--linear}) and also the reciprocal relations. Moreover, the entropy generation (\ref{entro--com233}) is zero when the thermodynamic equilibrium conditions are satisfied within the system (i.e. thermodynamic forces are zero at equilibrium).
\subsubsection{Galilean~invariance}
Another requirement which the entropy generation (\ref{entro--com233}) should satisfy is that it be invariant under a Galilei transformation \cite{BookGroot1985}, since the notions of reversible and irreversible behaviour must be invariant under such a transformation. It can be seen that the entropy generation (\ref{entro--com233}) satisfies automatically this requirement. Moreover, the model equations must be Galilean invariant as well, where, according to the classical mechanics, the balance equations must be the same in the inertia frames. It can be observed that our system equations satisfy this requirement. Note that, in another phase-field model (\cite{Abels2010}), the volume-averaged velocity is employed, which leads to a non-objective scalar term appearing in the chemical potential equation. Therefore a particular formulation for the convective terms is needed to keep the Galilean invariance of their model equations. In our model equations, on the other hand, the mass-averaged velocity is employed for the mixture, therefore no non-objective terms are involved. The system equations satisfy the Galilean invariance automatically.
\Section{Phase-field model for quasi-incompressible fluid with thermocapillary effects}
In this section, we develop a model of a binary Cahn-Hilliard fluids with thermocapillary effects in which both components are incompressible.
\subsection{Derivation of the model}
In order to study situations in which the density in each phase is uniform, it is convenient to adopt a thermodynamic formation which does not employ the density as an independent variable, as in the model of quasi-incompressible flow considered by \cite{Lowengrub1998}. Following their work, we choose the pressure and temperature as independent variables, and work with the Gibbs free energy. In addition, for a binary incompressible fluid system, the free energy density can appear as the “per unit mass” quantity or “per unit volume” quantity. In most phase-field models for two-phase flows (e.g. \cite{Hohenberg1977, LiuShen2002}), the density of two components are assumed to be constant and equal, and the per unit mass and per unit volume specification of the free energy density are equivalent. However, in the situation we study here, the densities of two fluids of the mixture may not be matched and thus the per unit mass and per unit volume forms are not equivalent. As we mentioned above, several models have been developed for the binary incompressible fluid with different densities, in which the per unit volume form of free energy density was employed in \cite{Boyer2002,Ding2007,shen2010,Abels2012} and the per unit mass form by \cite{Lowengrub1998}. Here we concentrate on the Gibbs free energy density in the per unit mass form, and denote it by $\hat{g}(T,p, c, \nabla c)$. Again, similar to the definition of the free energy (\ref{internal--free--energy}) for binary compressible fluid, we introduce the gradient terms (gradient energy) into the Gibbs free energy of our model, which can then be given in the form
\begin{eqnarray}
&\hat{g}(T,p, c, \nabla c)=g(T, p,  c)+g^{grad}(T,\nabla   c),~~~~g^{grad}=f^{grad}=\lambda_{f}(T)\frac{1}{2}|\nabla   c|^2,\label{internal--gibbsenergy}
\end{eqnarray}
where $g$ is the classical parts of the Gibbs free energy density, and $\lambda_f(T)$ is a temperature dependent coefficient and will lead to the thermocapillary effects along the interface (see $\S 4.3$ for details). For the classical part of the internal energy defined by (\ref{def--internal--com}), we have the following thermodynamic relation
\begin{eqnarray}
u(s, \rho,  c)=g(T, p,   c)+Ts-\frac{p}{\rho}.\label{cond--quasi--en}
\end{eqnarray}
Using the thermodynamic relation (\ref{com-der-e}) leads to
\begin{eqnarray}
{\rm d}g(T, p,   c)={\rm d}u(s, \rho,  c)-{\rm d}(sT)+{\rm d}\bigg(\frac{p}{\rho}\bigg)=-s{\rm d}T +\frac{1}{\rho}{\rm d}p +\frac{\partial u}{\partial   c} \bigg|_{\rho,s} {\rm d} c,\label{com-der-e1}
\end{eqnarray}
where we note the relations
\begin{eqnarray}
\frac{\partial g(T,p,  c)}{\partial T}\bigg|_{p,  c}=-s ,~~~~\frac{\partial g(T,p,  c)}{\partial p}\bigg|_{T,  c}=\frac{1}{\rho}~~~~{\rm and}~~~~\frac{\partial g(T,p,  c)}{\partial   c}\bigg|_{T,p}=\frac{\partial u}{\partial   c} \bigg|_{\rho,s}.\label{cond--quasi--free}
\end{eqnarray}
Here as we notice that the variable density is independent of temperature and pressure (See (2.5)), the condition of the incompressibility can then be written in the terms of the Gibbs free energy
\begin{eqnarray}
\frac{\partial^2 g(T,p,  c)}{\partial^2 p }=0,\label{cond--quasi--gibb}
\end{eqnarray}
where the second condition in (\ref{cond--quasi--free}) is used. Condition (\ref{cond--quasi--gibb}) implies that Gibbs free energy is a linear function of pressure, (e.g. \cite{Lowengrub1998})
\begin{eqnarray}
g(T,p,c)=g_0(T,c)+\frac{p}{\rho(c)}.\label{cond--gibbs}
\end{eqnarray}
We then re-define the classical internal energy as a function of $T$, $p$ and $c$,
\begin{eqnarray}
\tilde{u}(T, p,  c)=u(s,\rho,c)=g(T, p,   c)+Ts-\frac{p}{\rho(c)},\label{def--quasi--internal}
\end{eqnarray}
where the relations (\ref{com-der-e1}) and (\ref{cond--quasi--free}) still hold. Similarly to the definition of the internal energy (\ref{def--internal--com}) and entropy (\ref{internal--energy}) for the binary compressible fluid model, the specific internal energy $\hat{u}$ and the specific entropy $\hat{s}$ for binary incompressible fluids can be re-defined in the form
\begin{eqnarray}
&&\hat{u}(T,p, c, \nabla c)=\tilde{u}(T,p, c)+u^{grad}(\nabla   c),~~~~u^{grad}=\lambda _{u}\frac{1}{2}|\nabla   c|^2,\\
&&\hat{s}(T, c, \nabla c)=\tilde{s}(T,c)+s^{grad}(\nabla   c),~~~~~~~~~~~s^{grad}=\lambda_{s}\frac{1}{2}|\nabla   c|^2,\label{def--quasi--entropy}
\end{eqnarray}
where $\tilde{u}$ and $\tilde{s}$ are the classical parts of the specific internal energy and entropy associated with the Gibbs free energy, $\lambda_{u}$ and $\lambda_{s}$ are constant. In addition to these classical contributions, we assume that the same thermodynamics relations that hold for the classical terms also hold for the total terms, such that
\begin{eqnarray}
\hat{g}=\hat{u}-T\hat{s}+\frac{p}{\rho}=\hat{f}+\frac{p}{\rho},~~~~\frac{\partial \hat{g}}{\partial T}\bigg|_{p,  c, \nabla   c}=-\hat{s}.\label{helm--free--sub2}
\end{eqnarray}
Thus, from (\ref{internal--gibbsenergy}), (\ref{def--quasi--internal})-(\ref{def--quasi--entropy}), the relation for the coefficients (\ref{con--free--1}) holds as well. The specifications of these three coefficients will be discussed in $\S 4.2$. Note that $\lambda_{u}$, $\lambda_{s}$ together with $\lambda_{f}(T)$ (in Eq.(\ref{internal--gibbsenergy})) can be further used to relate the surface tension of phase-field model to that of sharp-interface model when our phase-field model reduces to its sharp-interface limit (see $\S 5.4$ for details).\\
Now we derive the system of equations for the quasi-incompressible phase-field model. We still use (\ref{vol--com--mass})-(\ref{vol--com--phi}) to define the total properties, namely mass $M$, momentum $\bold{P}$, energy $E$, entropy $S$ and mass constituent $C$ in a material control volume $V(t)$ of the domain $\Omega$. We further assume that the corresponding general balance laws (\ref{der--com--mass})-(\ref{der--com--phi}) that hold for the binary compressible fluid also hold for the quasi-incompressible fluid, which can then be written as
\begin{align}
\frac{{\rm D} \rho}{{\rm D} t} &= - \rho (\nabla \cdot \bold{v}),\label{equ--quasi--mass}\\
\rho \frac{{\rm D} \bold{v}}{{\rm D} t} &=\nabla \cdot \bold{m}-\rho g \hat{\bold{z}},\label{equ--quasi--mom}\\
\rho \frac{{\rm D} \tilde{u}}{{\rm D} t}&=-\nabla \cdot (  \rho \lambda_{u}\frac{{\rm D}   c}{{\rm D} t}\nabla   c) +\rho\lambda_{u} \Delta c \frac{{\rm D}    c}{{\rm D} t}  +(\bold{m} -\rho \lambda_{u}|\nabla c|^2\bold{I}+\rho \lambda_{u}\bold{T}) : \nabla \bold{v} -\nabla \cdot  \bold{q}_E \nonumber\\
&- \nabla \cdot \bold{q}^{nc}_E,\label{equ--quasi--en1}\\
 \rho\frac{{\rm D} \tilde{s}}{{\rm D} t}&=-\nabla \cdot (\rho \lambda_{s} \frac{{\rm D}   c}{{\rm D} t}\nabla   c) +\rho\lambda_{s}\Delta c \frac{{\rm D}    c}{{\rm D} t}-(\rho  \lambda_{s}|\nabla c|^2\bold{I}-\rho  \lambda_{s}\bold{T}) : \nabla \bold{v}-\nabla \cdot (\frac{\bold{q}_{E}}{T})-\nabla \cdot \bold{q}_S^{nc}\nonumber\\
 &+S_{gen}, \label{equ--quasi--entro}\\
\rho \frac{{\rm D}   c}{{\rm D} t}&=- \nabla \cdot \bold{q}_{C},\label{equ--quasi--phi}
\end{align}
where, as the pressure $p$ is not defined in the traditional way, the general stress tensor $\bold{m}$ is not defined explicitly.\\
Note that, in contrast to the case of binary compressible fluid, the classical part of internal energy $\tilde{u}$ we defined here does not depend on the entropy $\tilde{s}$ directly. However, as the derivations are carried out within the thermodynamic framework that is based on the entropy generation, a thermodynamic relation between the internal energy and entropy is still needed. Having in mind the definition of the internal energy (\ref{def--quasi--internal}) and using the relations (\ref{com-der-e1}) and (\ref{cond--quasi--free}), we obtain the following relation between the classical part of internal energy $\tilde{u}$, Gibbs free energy $g$ and entropy $\tilde{s}$
\begin{eqnarray}
\rho \frac{{\rm D} \tilde{u}}{{\rm D} t}= \rho \frac{\partial g_0}{\partial   c} \frac{{\rm D}   c}{{\rm D} t}+\rho T \frac{{\rm D} \tilde{s}(T,   c)}{{\rm D} t}.\label{equ--quasi--en}
\end{eqnarray}
Then similar to the method used for the binary compressible fluid model, we use the unknowns, including $\bold{m}$, $\bold{q}_{E}$, $\bold{q}^{nc}_{E}$, $\bold{q}^{nc}_{S}$ to express the entropy generation in the form
\begin{eqnarray}
&&S_{gen}=\frac{1}{T}\bigg(\bold{m}-\rho \lambda_{f}(T)|\nabla c|^2\bold{I}+\rho  \lambda_{f}( T)\bold{T}  \bigg) : \nabla \bold{v}+\nabla \frac{1}{T} \cdot \bigg(  \rho \lambda_{u}\frac{{\rm D}   c}{{\rm D} t}\nabla   c+ \bold{q}_E  +  \bold{q}^{nc}_E\nonumber\\
&&-\tilde{\mu}_{C}\bold{q}_{C}\bigg) -\nabla \cdot \bigg[\frac{1}{T}\bigg(\rho \lambda_{f}( T)  \frac{{\rm D}   c}{{\rm D} t}\nabla   c +\bold{q}^{nc}_E-\tilde{\mu}_{C} \bold{q}_{C} -T\bold{q}_S^{nc}\bigg)\bigg]- \frac{1}{T}\bold{q}_{C}\cdot\nabla  \tilde{\mu}_{C},\label{entropy11}
\end{eqnarray}
where we have used Eqs.(\ref{equ--quasi--mass})-(\ref{equ--quasi--en}). $\tilde{\mu}_{C}=\partial g_{0}(c)/\partial   c-\lambda_{f}(T)\Delta c$ is a potential term. As the pressure is no longer defined by the thermodynamic formulas in this model, we now derive the pressure in an alternative way that used in \cite{Lowengrub1998}, where the pressure was obtained from the non-dissipated part of the general stress $\bold{m}$. Considering a dissipative process, we denote the general stress tensor by $\bold{m}=\bold{m}_0+\boldsymbol{\tau}$, in which $\boldsymbol{\tau}$ is the deviatoric stress tensor with zero trace, and $\bold{m}_0$ is the unknown part to be defined later. We then denote $\bold{D}\bold{v} = \nabla\bold{v}-( \nabla \cdot\bold{v})\bold{I}/3$ as the deviatoric part of $\nabla \bold{v}$ (${\rm tr}~\bold{Dv}=0$). The entropy expression (\ref{entropy11}) can be rewritten as:
\begin{align}
S_{gen}&=\frac{1}{T}\bigg(\bold{m}_0-\rho \lambda_{f}(T)|\nabla c|^2\bold{I}+\rho \lambda_{f}(T)\bold{T}  \bigg) : \bold{Dv}+\frac{1}{T}\boldsymbol{\tau} :\nabla \bold{v}+\nabla \frac{1}{T} \cdot \bigg\{ \rho \lambda_{u}\frac{{\rm D}   c}{{\rm D} t}\nabla   c+\bold{q}_E\nonumber\\
  &+ \bold{q}^{nc}_E-\bigg(\tilde{\mu}_{C}+\big(\frac{1}{3}{\rm tr}~\bold{m}_{0}-\frac{2}{3}\rho  \lambda_{f}(T)|\nabla c|^2\big)\frac{1}{\rho^2}\frac{{\rm d} \rho}{{\rm d} c}\bigg)\bold{q}_{C}\bigg\}  -\nabla \cdot \bigg\{ \frac{1}{T}\bigg[\rho\lambda_{f}(T) \frac{{\rm D}   c}{{\rm D} t}\nabla   c \nonumber\\
&+\bold{q}^{nc}_E-\bigg(\tilde{\mu}_{C}+\big(\frac{1}{3}{\rm tr}~\bold{m}_{0}-\frac{2}{3}\rho  \lambda_{f}( T)|\nabla c|^2\big)\frac{1}{\rho^2}\frac{{\rm d} \rho}{{\rm d} c} \bigg) \bold{q}_{C} -T\bold{q}_S^{nc}\bigg]\bigg\}-\frac{1}{T}\bold{q}_{C}\cdot\nabla\bigg(\tilde{\mu}_{C} \nonumber\\
&+\big(\frac{1}{3}{\rm tr}~\bold{m}_{0}-\frac{2}{3}\rho  \lambda_{f}(T)|\nabla c|^2\big)\frac{1}{\rho^2}\frac{{\rm d} \rho}{{\rm d} c}\bigg),\label{entropyproduciton}
\end{align}
where we have used the mass balance (\ref{equ--quasi--mass}) and the following identity
\begin{eqnarray}
&&\bigg(\bold{m}_0-\rho \lambda_{f}(T)|\nabla c|^2\bold{I}+\rho \lambda_{f}(T)\bold{T}\bigg):\frac{1}{3}(\nabla \cdot \bold{v})\bold{I}\nonumber\\
&&=\frac{1}{3}\bigg({\rm tr}~\bold{m}_0-\rho \lambda_{f}(T){\rm tr}~|\nabla c|^2\bold{I}+\rho  \lambda_{f} (T) {\rm tr}~\bold{T}\bigg)(\nabla \cdot \bold{v})\nonumber\\
&&=\bigg(\frac{1}{3} {\rm tr}~\bold{m}_0-\frac{2}{3}\rho \lambda_{f}(T)|\nabla c|^2\bigg)(\nabla \cdot \bold{v}).
\end{eqnarray}
Now we assume that the first two terms on the right-hand side of (\ref{entropyproduciton}) are non-dissipative and define the pressure $p$ by
\begin{eqnarray}
-p=\frac{1}{3}{\rm tr}~\bold{m}=\frac{1}{3}{\rm tr}\bigg(\bold{m}_0-\rho \lambda_{f}(T)|\nabla c|^2\bold{I}+\rho \lambda_{f}(T)\bold{T}\bigg)=\frac{1}{3}{\rm tr}~\bold{m}_0-\frac{2}{3}\rho \lambda_{f}(T)|\nabla c|^2,\label{pressure-ns}
\end{eqnarray}
such that
\begin{eqnarray}
-p\bold{I}=\bold{m}_0-\rho \lambda_{f}(T)|\nabla c|^2\bold{I}+\rho \lambda_{f}(T)\bold{T},
\end{eqnarray}
in which the way we use to define the pressure in Eq.(\ref{pressure-ns}) is analogous to the way that defines the kinematic pressure for the classical Navier-Stokes equations \cite{Batchelor2000}. To ensure our model is consistent with the second law of thermodynamics ($S_{gen}\geqslant 0$), we specify the unknown terms as the following
\begin{eqnarray}
&&\bold{q}_E=-k(c)\nabla T, \hspace{37mm}\bold{q}_{E}^{nc}=-\rho \lambda_{u}\frac{{\rm D}  c}{{\rm D} t}+\mu_{C}\bold{q}_{C},\label{energgg1}\\
&&\bold{q}_{S}^{nc}=-\rho \lambda_{s}\frac{{\rm D}   c}{ {\rm D} t} \nabla   c, \hspace{11mm}\bold{m}_0=-p\bold{I}+\rho \lambda_{f}(T)|\nabla c|^2\bold{I}-\rho \lambda_{f}( T)\bold{T},\label{energgg2}\\
&&\mu_{C} =\frac{\partial g_0}{\partial   c}-\frac{p}{\rho^2}\frac{{\rm d} \rho}{{\rm d} c}-\lambda_{f}(T) \Delta c, \hspace{26mm}\bold{q}_{C}=-m_C\nabla \mu_{C},\\
&&\bold{T} =\nabla c\otimes \nabla c, \hspace{19mm}\boldsymbol{\tau}=\mu(c)(\nabla \bold{v} +\nabla \bold{v}^{T} )-\frac{2}{3}\mu(c) (\nabla \cdot \bold{v})\bold{I}.\label{energgg3}
\end{eqnarray}
Here $\bold{m}=\bold{m}_0+\boldsymbol{\tau}$. Note that, comparing to the model developed by Lowengrub \cite{Lowengrub1998}, an extra term, $\rho \lambda_{f}(T)|\nabla c|^2\bold{I}$, appears in the stress tensor $\bold{m}_0$ (Eq.(\ref{energgg2})). This term could be absorbed into the pressure for convenience (see, \cite{Hua2011, LiuShen2002} for examples). However, we still keep this term in order that the surface gradient of the surface tension in the sharp interface limit can be recovered (See $\S 5.4$ for details). This treatment is similar to that used in \cite{Guo2013} and \cite{Garcke2014}, where in \cite{Garcke2014} this term is also required to recover the surface gradient term in the asymptotic analysis. Besides the momentum equation, the pressure appears in the chemical potential equation as well, which is different with the chemical potential (\ref{sys--com--phi2}) for the binary compressible fluid model. By substituting the above terms into Eqs.(\ref{equ--quasi--mass})-(\ref{equ--quasi--phi}), we obtain the system of equations for the phase-field model governing the quasi-incompressible fluid with thermocapillary effects
\begin{align}
\frac{{\rm D} \rho}{{\rm D} t} &= - \rho (\nabla \cdot \bold{v}),\label{sys--quasi--mass}\\
\rho \frac{{\rm D} \bold{v}}{{\rm D} t}&=\nabla \cdot \bold{m}-\rho g \hat{\bold{z}},\label{sys--quasi--mom}\\
\rho \frac{{\rm D} \tilde{u}}{{\rm D} t}&= \lambda_{u}\nabla \cdot ( \rho \nabla  c )\frac{{\rm D}    c}{{\rm D} t} +(-p\bold{I}-\rho T\lambda_{s}|\nabla c|^2\bold{I}+\rho T\lambda_{s}\bold{T}+\boldsymbol{\tau}) : \nabla \bold{v}+\nabla \cdot  (k(c)\nabla T\nonumber\\
&+m_{C}\mu_{C}\nabla \mu _{C}),\label{sys--quasi--en}\\
\rho \frac{{\rm D} \tilde{s}}{{\rm D} t}&= \lambda_{s}\nabla \cdot ( \rho \nabla  c )\frac{{\rm D}    c}{{\rm D} t} +\frac{1}{T}( -\rho T\lambda_{s}|\nabla c|^2\bold{I}+\rho T\lambda_{s}\bold{T}+\boldsymbol{\tau}) : \nabla \bold{v}+ \frac{1}{T}\nabla \cdot (k(c)\nabla T),\nonumber\\
&\label{sys--quasi--entro}\\
\rho \frac{{\rm D}   c}{{\rm D} t}&=m_C\Delta \mu_{C},\label{sys--quasi--phi1}\\
\mu_{C} &=\frac{\partial g_0}{\partial   c}-\frac{p}{\rho^2}\frac{{\rm d} \rho}{{\rm d} c}-\lambda_{f}(T) \Delta   c. \label{sys--quasi--phi2}
\end{align}
Multiplying Eqs.(\ref{sys--quasi--mass}), (\ref{sys--quasi--mom}) and (\ref{sys--quasi--entro})-(\ref{sys--quasi--phi2}) by $p/\rho+\bold{v}\cdot \bold{v}/2+u$, $\bold{v}$, $T$, $\mu_{C}$ and $\rho {\rm D}  c/{\rm D}t$, and summing them up, we can obtain the first law of thermodynamics (\ref{der--com--en}) that we used to derive the model. By substituting the terms, including $\bold{m}$, $\bold{q}_{E}$, $\bold{q}^{nc}_{E}$, $\bold{q}^{nc}_{S}$ and $\bold{q}_{C}$ into the entropy generation (\ref{entropyproduciton}), we obtain the second law of thermodynamics for our phase-field model,
\begin{eqnarray}
S_{gen} =\frac{1}{T}\boldsymbol{\tau} : \nabla \bold{v} + k(c)|\frac{\nabla T}{T}|^2+\frac{m_C}{T}|\nabla \mu_{C}|^2 \geqslant 0.\label{entro--quasi}
\end{eqnarray}
Similar to the binary compressible model, the choices of the terms $\boldsymbol{\tau}, \bold{q}_E$ and $\bold{q}_C$ satisfy the linear relation (\ref{relation--linear}) and the Onsager reciprocal relations ($\S 3.2.2$). Moreover, it can be observed that the entropy generation (\ref{entro--quasi}) and the system equations are Galilean invariant.\\
As mentioned above, several phase-field models have been developed for two-phase flows with thermocapillary effects. However, in most of these models, the classical energy balance equation
\begin{eqnarray}
\rho c_{hc}\frac{{\rm D} T}{{\rm D} t}= \nabla \cdot  (k\nabla T),\label{classical--energy}
\end{eqnarray}
was incorporated directly into the phase-field model, where thermodynamic consistency can be hardly maintained. Comparing with the classical energy balance equation (\ref{classical--energy}), several extra terms appear in our energy balance equation (\ref{sys--quasi--en}), which guarantee thermodynamic consistency (see $\S 4.2$).\\
Note that, if we define a new pressure as $\bar{p}=p-\rho \lambda_{f}(T)|\nabla c|^2$, and substitute it into the system equations (\ref{sys--quasi--mass})-(\ref{sys--quasi--phi2}), our model, in the isothermal case, reduces to the quasi-incompressible NSCH model developed by \cite{Lowengrub1998}.\\
By using the variable mass density (\ref{variable--c}), the mass balance equation (\ref{sys--quasi--mass}) can be further rewritten as
\begin{eqnarray}
\nabla \cdot \bold{v}=-\frac{1}{\rho}\frac{{\rm D}\rho}{{\rm D} t}=\alpha\rho\frac{{\rm D}c}{{\rm D} t}=\alpha  m_C\Delta \mu_C,\label{mumu11}
\end{eqnarray}
where we have used Cahn-Hilliard equation (\ref{sys--quasi--phi1}) and let $\alpha=(\rho_2-\rho_1)/\rho_2\rho_1$.\\
The initial conditions are given by
\begin{eqnarray}
\bold{v}|_{t=0}=\bold{v}_0,~~~~c|_{t=0}=c_0,~~~~{\rm and}~~~~T|_{t=0}=T_0.
\end{eqnarray}
For the velocity, the usual no-slip boundary conditions can be posed on $\partial \Omega$
\begin{eqnarray}
\bold{v}=\bold{v_b}.
\end{eqnarray}
For the phase field, it is normal to employ Neumann boundary conditions on $\partial \Omega$
\begin{eqnarray}
\nabla c\cdot \hat{\bold{n}}=h_c,~~~~{\rm and}~~~~\nabla\mu_C\cdot \hat{\bold{n}}=h_{\mu}.
\end{eqnarray}
For the temperature, Dirichlet and Neumann boundary conditions can be posed on $\partial \Omega$
\begin{eqnarray}
T=T_b,~~~~{\rm or}~~~~\nabla T\cdot \hat{\bold{n}}=q_b
\end{eqnarray}
for the specified temperature and heat flux on the boundary $\partial \Omega$ respectively, and Robin boundary conditions can be posed as well.
\subsection{Specifications of the model}
We now specify the properties including the Gibbs free energy, entropy and internal energy for our phase-field model (\ref{sys--quasi--mass})-(\ref{sys--quasi--phi2}). In \cite{Anderson2000}, a phase-field model for the solidification of a pure material that includes convection in the liquid phase was developed, in which the case of the quasi-incompressibility (assuming that the density in each phase is uniform) was discussed. In their work, the Gibbs free energy was suggested in the form
\begin{eqnarray}
&&\hat{g}(T,p, c, \nabla c)=g_0(T,  c)+\frac{p}{\rho(  c)}+\lambda_{f}(T)\frac{1}{2}|\nabla   c|^2,\label{gibbsfree}\\
&&g_0(T,  c)=\big(u_0-c_{hc}T_0 \big)(1-\frac{T}{T_0})-c_{hc}T {\rm ln}(\frac{T}{T_0})+\gamma_f(T) h(c),\label{gibbsfree--22}
\end{eqnarray}
which we have adopted for the present work. Here, $c_{hc}$ is the heat capacity, $T_0$ is the reference temperature, $\tilde{u}_0$ is the reference internal energy corresponding to $T_0$, and $\gamma(T)$ is a temperature dependent parameter that will be discussed later in this section. The free energy function $h(c)$ is a double-well potential and is given by
\begin{eqnarray}
h(c)=\frac{c^2(c-1)^2}{4},\label{free--energy--hc}
\end{eqnarray}
where the wells define the phases, and lead to an interfacial layer with large variations for $c$ (e.g. \cite{Gurtin1996}). Note that the form for $\hat{g}$ (\ref{gibbsfree}) is consistent with the incompressible condition (\ref{cond--gibbs}), which is a linear function of pressure. Moreover, this form for $\hat{g}$ is consistent with an internal energy $\hat{u}$, which is a linear function of temperature and leads to the classical heat equation in the bulk liquid \cite{Wang1993, Anderson2001}. The corresponding expressions for the entropy and internal energy are assumed in the form
\begin{align}
\hat{s}&=\tilde{s}+s^{nc}=\frac{1}{T_0} u_0 +c _{hc} {\rm ln}(\frac{T}{T_0})+\gamma_s h(c)+\lambda_{s}\frac{1}{2}|\nabla   c|^2,\\
\hat{u}&=\tilde{u}+u^{nc}=\tilde{u}_0+c_{hc}(T- T_0)+\gamma _u h(c)+\lambda_{u}\frac{1}{2}|\nabla   c|^2,\label{specified--internal--energy1}
\end{align}
where $\tilde{u}_0$ is the reference internal energy corresponds to $T_0$.\\
We now specify those coefficients, including $\lambda_f(T)$, $\lambda_s$, $\lambda_u$, $\gamma_f(T)$, $\gamma_s$ and $\gamma_u$, which are used to define the internal energy, entropy and free energy of the system (Eqs.(\ref{gibbsfree})-(\ref{specified--internal--energy1})). In the sharp-interface model for the thermocapillary flow, the interface is usually represented as a surface of zero thickness with the surface tension as its physical property. An equation of state is required to relate the surface tension to the temperature, where for the sake of simplicity, we only consider a linear relation in this study,
\begin{eqnarray}
&&\sigma(T)=\sigma_0-\sigma_T(T-T_0),\label{surface--tension--sharpinterface}
\end{eqnarray}
where $\sigma_0$ is the interfacial tension at the reference temperature $T_0$, $\sigma_T$ is the rate of change of interfacial tension with temperature, defined as $\sigma_T = \partial\sigma(T)/\partial T$.  In our phase-field model, however, the interface has finite thickness and the extra reactive stress (Ericksen's stress) $\bold{T}$ (Eq.(\ref{energgg3})) appears in Navier-Stokes equation to mimic the surface tension, where the coefficient of $\bold{T}$,
\begin{eqnarray}
&&\lambda_f(T)=\lambda_u-T\lambda_s,\label{surface--tension--diffuseinterface}
\end{eqnarray}
is a linear function of temperature. We then try to relate $\sigma(T)$ and $\lambda(T)$ by introducing two parameters: the first parameter is $\epsilon$ with respect to the diffuse interface thickness, and the second one $\eta$, a ratio parameter that relates the two surface tensions. As the interface thickness goes to zero, our phase-field model reduces to its sharp-interface limit, and the value of $\eta$ can then be determined (see $\S 5.4$ for details). The corresponding coefficients can then be given as
\begin{eqnarray}
&&\lambda_f(T)=\eta\epsilon\sigma(T)=\eta\epsilon\sigma_0-\eta\epsilon\sigma_T(T-T_0),~~~~\gamma_f(T)=\frac{\eta}{\epsilon}\sigma(T)=\frac{\eta}{\epsilon}\sigma_0-\frac{\eta}{\epsilon}\sigma_T(T-T_0),\nonumber\\
&&\lambda_s=\eta\epsilon\sigma_T,\hspace{50mm}\gamma_s=\frac{\eta}{\epsilon}\sigma_T,\nonumber\\
&&\lambda_u=\eta\epsilon\sigma_0+\eta\epsilon\sigma_T T_0,\hspace{35mm}\gamma_u=\frac{\eta}{\epsilon}\sigma_0+\frac{\eta}{\epsilon}\sigma_T T_0.\label{epsilon--condition11}
\end{eqnarray}
Here the coefficients $\lambda_f(T)$, $\lambda_s$ and $\lambda_u$ for the gradient terms are of $O(\epsilon^2)$ of those coefficients $\gamma_f(T)$, $\gamma_s$ and $\gamma_u$ for the corresponding classical terms, which agrees with the definition of the Cahn-Hilliard free energy (e.g. \cite {Lowengrub1998, LiuShen2002}). With the specifications above, the total energy $E$ of our phase-field model can now be written as
\begin{eqnarray}
&&E =\int_{V(t)}\bigg(\frac{1}{2} \rho | \bold{v} |^2+ \rho g  z +\rho \tilde{u}_0+ \rho c_{hc}(T- T_0)+\rho\gamma _u h(c)+\rho\lambda_{u}\frac{1}{2}|\nabla   c|^2 \bigg){\rm d}V.\nonumber
\end{eqnarray}
\subsection{Non-dimensionalization}
With the help of the specification in Eq.(\ref{epsilon--condition11}), we non-dimensionalize the phase-field model (\ref{sys--quasi--mass})-(\ref{sys--quasi--en}), (\ref{sys--quasi--phi1}) and (\ref{sys--quasi--phi2}) as follows: we let $L^{\star}$, $V^{\star}$ and $T^{\star}$ denote the characteric scales of length, velocity and temperature. Then introduce the dimensionless variables $\bar{\bold{x}}=\bold{x}/L^{\star}$, $\bar{t}=V^{\star}t/L^{\star}$, and also $\bar{\epsilon}=\epsilon/ L^{\star}$, $\bar{\bold{v}}=\bold{v}/V^{\star}$, $\bar{p}=p\rho_{1}\mu_{C}^{\star}$, $\bar{\mu}_{C}=\mu_{C}/\mu^{\star}_{C}$. For the variable density $\rho(c)$, viscosity $\mu(c)$ and thermal conductivity $k(c)$ (Eqs.(\ref{variable--c}) and (\ref{variable--thermaldiffusicity})), we let $\rho_{i}$, $\mu_{i}$ and $k_{i}$ (i=1,2) denote the corresponding properties of the $i$th fluid, and introduce the dimensionless variables $\bar{\rho}_r=\rho(c)/\rho_{1}$, $\bar{\mu}_r=\mu(c)/\mu_{1}$ and $\bar{k}_r=k(c)/k_{1}$. Moreover, for the temperature field, we introduce a new dimensionless variable $\bar{T}=(T-T_{0})/T^{\star}$. The surface tension $\sigma(T)$ (Eq.(\ref{surface--tension--sharpinterface})) is scaled by $\sigma_{0}$ such that $\bar{\sigma}(T)=\sigma(T)/\sigma_{0}$. $\sigma_{T}$ is then scaled by $\sigma_{0}/T^{\star}$, such that $\bar{\sigma_{T}}=\sigma_{T}T^{\star}/\sigma_{0}$. Omitting the bar notation, our phase-field model can now be rewritten as
\begin{align}
\nabla \cdot \bold{v}&=-\frac{1}{\rho_r}\frac{\partial\rho_r}{\partial c} \frac{{\rm D} c}{{\rm D} t},\label{non--sys--quasi--mass}\\
\rho_r\frac{{\rm D} \bold{v}}{{\rm D} t}&=-\frac{1}{M}\big[\nabla p-Ca\nabla \big(\rho_r \sigma(T)|\nabla c|^2\big) +{Ca}\nabla\cdot \big(\rho_r \sigma(T)\bold{T}\big)\big]\nonumber\\
&+\frac{1}{Re}\nabla\cdot \big(\mu_r(\nabla \bold{v} +\nabla \bold{v}^{T} )-\frac{2}{3}\mu_r (\nabla \cdot \bold{v})\bold{I}\big)- \frac{\rho_r}{ Fr^2} \hat{\bold{z}} ,\label{non--sys--quasi--mom}\\
\rho \frac{{\rm D}   c}{{\rm D} t}&=\frac{1}{Pe}\Delta \mu_{C},\label{non--sys--quasi--phi1}\\
\mu_{C}&=\frac{Ca}{\epsilon^2}\sigma(T)\frac{{\rm d} h(c)}{{\rm d} c}-\frac{p}{\rho_r^2}\frac{{\rm d}\rho_r}{{\rm d} c}-\frac{Ca}{\rho_r}\sigma(T)\Delta  c, \label{non--sys--quasi--phi2}\\
\rho_r c_{hc} \frac{{\rm D} T}{{\rm D} t}&=\frac{EcCa}{M}(1+\sigma_T T_0)\nabla \cdot ( \rho_r \nabla  c )\frac{{\rm D}    c}{{\rm D} t} +\frac{Ec}{M}(-p\bold{I} -{Ca}\rho_r T \sigma_{T}|\nabla c|^2\bold{I} +{Ca}\rho_r T \sigma_{T} \bold{T}\nonumber\\
&+\frac{1 }{Re}\boldsymbol{\tau} ) : \nabla \bold{v}+\nabla \cdot  \big(\frac{1}{Ma}k_r\nabla T+\frac{Ec}{MPe}\mu_{C}\nabla \mu _{C}\big),\label{non--sys--quasi--en}
\end{align}
where $M={V^2}/{ \mu_{C}}$ is an analogue of the Mach number, $Ca={\eta\epsilon \sigma_{0}}/{ \mu_{C}L}$ is the Capillary number that measures the thickness of the interface, $Re={\mu_{1}}/{\rho_{1}VL}$ is the Reynolds number, $Fr= { V^2}/{gL}$ is the Froude number, $Pe={\rho LV}/{m_C\mu_{C}}$ is the diffusional Peclet number, $Ec={V^2}/{c_{ch}T}$ is the Eckert number that characterizes energy dissipation, and $Ma={\rho c_{ch}VL}/{k_{1}}$ is the Marangoni number. Note that this non-dimensional system equations will be computed to study the effects of Marangoni number through the example of thermocapillary migration of a drop. See $\S 6.5$ for details.
\Section{Sharp-interface limits}
Theoretically, there are usually two ways to validate the phase-field model. The first, as mentioned above, is to show thermodynamic consistency of the model. The second is to relate the phase-field model to its sharp-interface counterpart. Based on the assumption that a given sharp-interface formulation is the correct description of the physics under consideration, the phase-field model can be justified by simply showing that it is asymptotic to the classical sharp-interface description. In the isothermal case, some sharp-interface limit analyses have been carried out for the phase-field model of two-phase flow to show that the corresponding sharp-interface equations and jump conditions across the interface can be recovered from the phase-field model (e.g. \cite{Lowengrub1998,XPWang2007,Abels2012}). However, much less attention has been paid on the asymptotic analysis of the phase-field model for two-phase flows in the non-isothermal case, (e.g. thermocapillary flows, solidifications). \cite{Antanovskii1995} presented a phase-field model to study the thermocapillary flow, and showed that the hydrostatic equilibrium condition for the case of a flat interface and the Laplace-Young condition for the case of a drop in equilibrium can be recovered from his phase-field model. \cite{Jasnow1996} extended Model-H to study the thermocapillary flow, including the migration of a drop and spinodal decomposition of a binary fluid under a constant temperature gradient. In the corresponding sharp-interface limit, they showed that the additional stress term in the Navier-Stokes equation of their phase-field model is equivalent to the tangential and normal force of the appropriate sharp-interface model. \cite{Anderson2000} developed a phase-field model of solidification with convection in the melt, in which the two phases are considered as viscous liquids. In the sharp-interface analysis \cite{Anderson2001}, they used the matched asymptotic expansions to show that the standard boundary conditions, including Young-Laplace and Stefan conditions can be recovered from their phase-field model.
\subsection{Pillbox argument}
\begin{figure}
  \centerline{\includegraphics[width=0.65\textwidth]{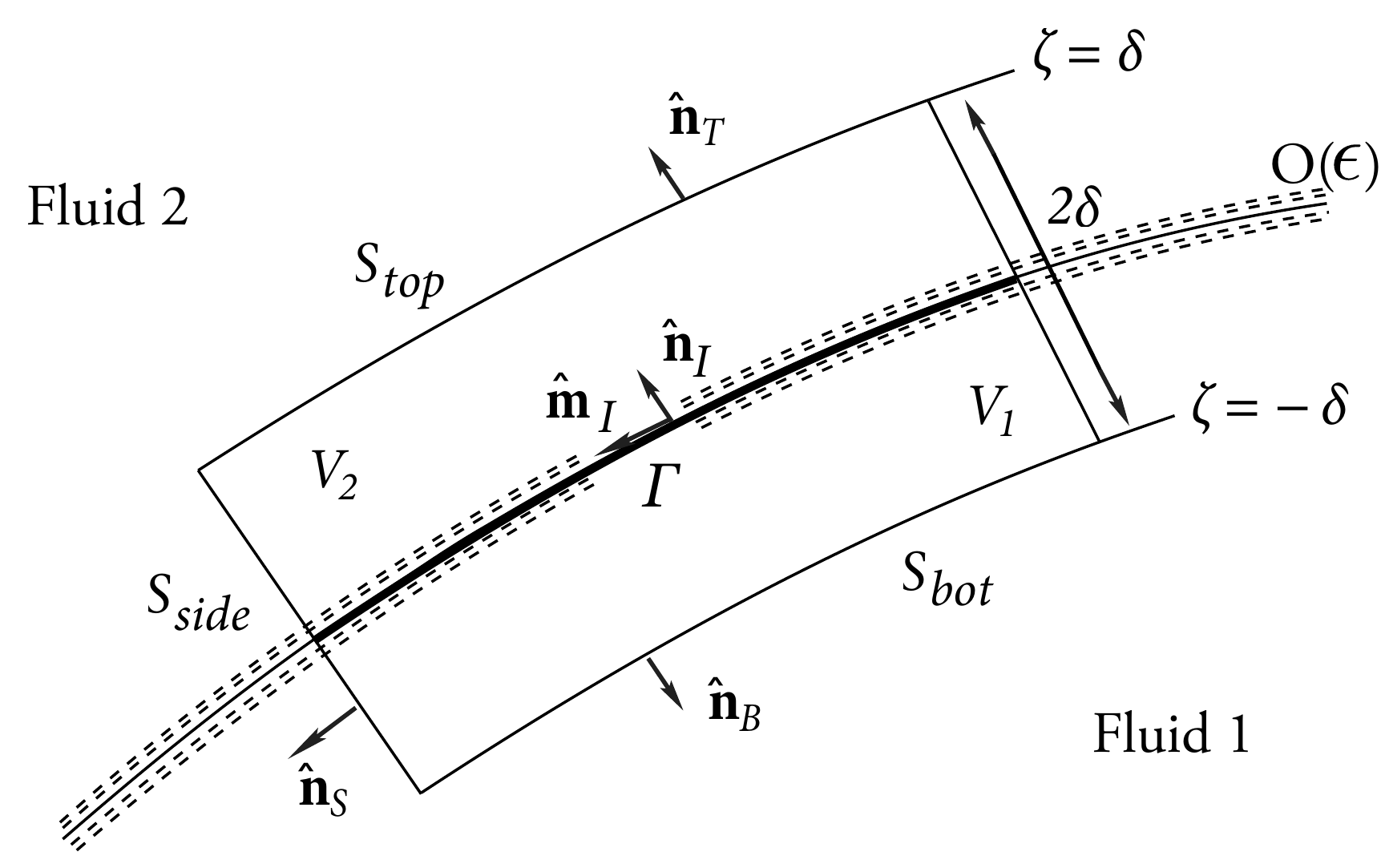}}
  \caption{A schematic diagram showing a diffuse interface between two fluids intersecting with a pillbox shaped control volume. $\hat{\bold{n}}_T$, $\hat{\bold{n}}_B$ and $\hat{\bold{n}}_S$ stand for the unit normal vector of the pillbox boundary on its top, bottom and side, respectively. The dotted lines represent the diffuse interface with thickness $O(\epsilon)$. $2\delta$ is the thickness of the pillbox. In the limit $\epsilon\ll \delta \ll L$, the interface thickness goes to zero, and the interface has constant density. $\hat{\bold{n}}_{I}$ and $\hat{\bold{m}}_{I}$ stand for the unit normal and tangent vector of the interface.}\label{sharpinterface2--box}
\end{figure}
In this section, we apply a pillbox argument to our phase-field model (\ref{sys--quasi--mass})-(\ref{sys--quasi--phi2}). In contrast to sharp-interface model, the interface of the phase-field model is diffusive with finite thickness $O(\epsilon)$. The phase variable (here is mass concentration $c$) is chosen to characterize the different phases, which takes distinct values (here $c=0, 1$) for the different phases, and changes rapidly through the interfacial region. Within this interfacial region, we chose a contour line of $c$ (here $c=0.5$) to represent the dividing surface $\Gamma$ for the following derivations (See \cite{Gibbs1928, Everett1972, Rowlinson1982} for details of the dividing surface). Moreover, as the largest variations of the phase variable occur in the direction normal to the interface, the side faces of the pillbox need to be treated carefully. Figure \ref{sharpinterface2--box} shows the pillbox shaped control volume designed for our phase-field model, where the surface is divided into three parts, namely the top $S_{top}$, bottom $S_{bot}$ and side $S_{side}$ surfaces with their unit normal vector $\hat{\bold{n}}_T$, $\hat{\bold{n}}_B$ and $\hat{\bold{n}}_S$ respectively. $V=V_{1}+V_{2}$ is the volume of the pillbox, where $V_{i}$ is the volume of single component. The pillbox has thickness of $2\delta$, where the top of the pillbox is above the dividing surface $\Gamma$ at a height $\zeta=\delta$ and the bottom is below $\Gamma$ at a height $\zeta=-\delta$. Here, $\zeta$ is a local coordinate normal to the interface $\Gamma$. In addition, the pillbox contains a portion of the diffuse interface with thickness $O(\epsilon)$, in which $\Gamma$ stands for the dividing surface with its normal and tangent unit vector $\hat{\bold{n}}_I$ and $\hat{\bold{m}}_I$. The key limit in the pillbox argument is that $\epsilon\ll\delta\ll L$, where $L$ is a length scale associated with the outer flow. In this limit, the volume of the pillbox becomes negligible on the outer scales, but the variations in the concentration variable $c$ that define the interfacial region, occur over a region fully contained within the pillbox. Also in this limit, the top ($S_{top}$) and bottom surface ($S_{bot}$) of the pillbox collapse onto the interface $\Gamma$, and have the normal vectors with opposite directions
 \begin{eqnarray}
&&S_{top}=S_{bot}=\Gamma,~~~~\hat{\bold{n}}_T=\hat{\bold{n}}_I,~~~~\hat{\bold{n}}_B=-\hat{\bold{n}}_I~~~~{\rm and}~~~~\hat{\bold{n}}_S=\hat{\bold{m}}_I.\label{norm--condition}
\end{eqnarray}
Moreover, we assume that the dividing interface is moving with the velocity $\bold{v}_{I}$\cite{Anderson1996, Anderson1998}.
\subsection{Governing equations in sharp-interface limit}
We first derive the system of equations in bulk regions away from the interfacial region. Here we only concentrate on the equations of mass, momentum and energy balance. The system of equations (\ref{sys--quasi--mass})-(\ref{sys--quasi--en}) reduce to the classical equations appropriate for the incompressible flows in bulk regions 
\begin{align}
\nabla \cdot \bold{v}&=0,\label{limit--mass}\\
\rho_i \frac{{\rm D} \bold{v}}{{\rm D} t}&=-\nabla p+\nabla \cdot\big(\mu_i(\nabla \bold{v}+\nabla \bold{v}^{T})\big)-\rho_i g \hat{\bold{z}},\label{limit--mom}\\
\rho_i {c_{hc}}\frac{{\rm D} T}{{\rm D} t}&=\nabla \cdot  (k_i\nabla T)+\mu_i(\nabla \bold{v}+\nabla \bold{v}^{T}) : \nabla \bold{v},\label{limit--en}
\end{align}
where $\rho_i$, $\mu_i$ and $k_i$ are the corresponding physical properties for the $i$th fluid. We now seek to derive the jump conditions for Eqs.(\ref{limit--mass})-(\ref{limit--en}) at the interface from our phase-field model (\ref{sys--quasi--mass})-(\ref{sys--quasi--phi2}).
\subsection{Jump condition for mass balance}
In the limit $\epsilon\ll \delta\ll L$, we have the properties \cite{Anderson1996, Anderson1998}
\begin{align}
\int_{V}\frac{\partial \rho}{\partial t}{\rm d}V&\sim-\int_{S}\rho\bold{v}_{I}\cdot \bold{\hat{n}}{\rm d}S,\label{jump--mass51121}\\
\int_{V}\frac{\partial (\rho\bold{v})}{\partial t}{\rm d}V&\sim-\int_{S}\rho\bold{v}\otimes\bold{v}_{I}\cdot \bold{\hat{n}}{\rm d}S.\label{jump--mass51122}
\end{align}
Substituting Eq.(\ref{jump--mass51121}) into the integral of Eq.(\ref{equ--quasi--mass}), and using the divergence theorem we obtain
\begin{eqnarray}
\int_{S}\rho(\bold{v}-\bold{v}_{I})\cdot \bold{\hat{n}}{\rm d}S=0.\label{jump--mass5}
\end{eqnarray}
According to our pillbox argument, we break up the above surface integral into pieces for the top, bottom and side surfaces to obtain
\begin{eqnarray}
&&\int_{S_{top}} \rho(\bold{v}-\bold{v}_{I}) \cdot \bold{\hat{n}}_{T}{\rm d}S+\int_{S_{bot}} \rho(\bold{v}-\bold{v}_{I}) \cdot \bold{\hat{n}}_{B}{\rm d}S+\oint_{C}\int_{-\delta}^{\delta} \rho(\bold{v}-\bold{v}_{I})  \cdot \bold{\hat{n}}_{S}{\rm d}\zeta{\rm d}l=0.\nonumber\\
&&\hspace{100mm}\label{jump--mass611}
\end{eqnarray}
Here the surface integral of side portion is further written in term of a line integral on the surface and an integral in the normal direction $\hat{\bold{n}}_S$, where the line is a closed curve at the side of the control volume that parallel to the interface. For viscous fluid under normal operating conditions, it is an experimentally observed fact (like the no-slip boundary conditions at solid walls) that no slip takes place at the interface \cite{Tyggvason2011}. Therefore, in the limit $\epsilon\ll\delta\ll L$, we have 
\begin{eqnarray}
\bold{v}\cdot \bold{\hat{m}}_{I}\sim\bold{v}_{I}\cdot \bold{\hat{m}}_{I}.\label{sidecond}
\end{eqnarray}
This condition implies that the third left term in Eq.(\ref{jump--mass611}) is bounded and does not contribute to the integral. Eq.(\ref{jump--mass611}) can be reduced to
\begin{eqnarray}
\int_{\Gamma}\big[\rho(\bold{v}-\bold{v}_{I})\big]\cdot \bold{\hat{n}}_{I}{\rm d}S=0,\label{jump--mass6}
\end{eqnarray}
where $[\chi ]=\chi_2-\chi_1$ refers to the jump of the quantity $\chi$ across the singular interface. Since the pillbox control volume $V$ that contains a portion of the diffuse interface is arbitrary, the integrand in Eq.(\ref{jump--mass6}) must be zero. This then yields the mass balance jump condition at the interface in a two-phase fluid system
\begin{eqnarray}
\big[\rho(\bold{v}-\bold{v}_{I})\big]\cdot \bold{\hat{n}}_{I}=0.\label{jump-mass777}
\end{eqnarray}
Further if we assume that there is no phase change (i.e. no flux) across the interface, Eq.(\ref{jump-mass777}) reduces to the jump condition that
\begin{eqnarray}
\big[\bold{v}\big]\cdot\bold{\hat{n}}_I=0.
\end{eqnarray}
\subsection{Jump condition for momentum balance}
Substituting Eq.(\ref{jump--mass51122}) into the integral of momentum equation (\ref{sys--quasi--mom}), we obtain
\begin{eqnarray}
&&\int_{S}\bigg(\rho\bold{v}\otimes( \bold{v}-\bold{v}_{I})+p\bold{I}-\eta\epsilon\sigma(T)\rho|\nabla c|^2\bold{I}+\eta\epsilon\sigma(T)\rho\bold{T} -\mu(\nabla \bold{v} +\nabla \bold{v}^{T} )\nonumber\\
&&+\frac{2}{3}\mu (\nabla \cdot \bold{v})\bold{I} \bigg)\cdot \bold{\hat{n}}{\rm d}S=0,
\end{eqnarray}
where we have used the mass balance equation (\ref{sys--quasi--mass}), such that
\begin{eqnarray}
\rho\frac{{\rm D} \bold{v}}{{\rm D} t}=\rho\frac{{\rm D} \bold{v}}{{\rm D} t}+\big(\frac{{\rm D} \rho}{{\rm D} t}+\rho(\nabla \cdot \bold{v})\big)\bold{v}=\frac{\partial (\rho\bold{v} )}{\partial t}+\nabla \cdot (\rho\bold{v}\otimes \bold{v}).
\end{eqnarray}
Moreover, in the limit $\epsilon\ll\delta\ll L$, we assume that the gravitational term $\rho g \bold{\hat{z}}$ is bounded and thus does not contribute to the volume integral. We then break up the above surface integral into pieces for the top, bottom and sides of the pillbox to obtain
\begin{eqnarray}
&&\int_{S_{top}}\bigg(\rho\bold{v}\otimes (\bold{v}-\bold{v}_{I})+p\bold{I}-\eta\epsilon\sigma(T)\rho|\nabla c|^2\bold{I}+\eta\epsilon\sigma(T)\rho\bold{T} -\mu(\nabla \bold{v} +\nabla \bold{v}^{T} )\nonumber\\
&&+\frac{2}{3}\mu (\nabla \cdot \bold{v})\bold{I}\bigg)\cdot \bold{\hat{n}}_{T}{\rm d}S\nonumber\\
&&+\int_{S_{bot}}\bigg(\rho\bold{v}\otimes (\bold{v}-\bold{v}_{I})+p\bold{I}-\eta\epsilon\sigma(T)\rho|\nabla c|^2\bold{I}+\eta\epsilon\sigma(T)\rho\bold{T} -\mu(\nabla \bold{v} +\nabla \bold{v}^{T} )\nonumber\\
&&+\frac{2}{3}\mu (\nabla \cdot \bold{v})\bold{I}\bigg)\cdot \bold{\hat{n}}_{B}{\rm d}S\nonumber\\
&&+\oint_{C}\int_{-\delta}^{\delta} \bigg(\rho\bold{v}\otimes (\bold{v}-\bold{v}_{I})+p\bold{I}-\eta\epsilon\sigma(T)\rho|\nabla c|^2\bold{I}+\eta\epsilon\sigma(T)\rho\bold{T} -\mu(\nabla \bold{v} +\nabla \bold{v}^{T} )\nonumber\\
&&+\frac{2}{3}\mu (\nabla \cdot \bold{v})\bold{I}\bigg)\cdot \bold{\hat{n}}_S{\rm d}\zeta{\rm d}l=0.\label{jump-momen-11}
\end{eqnarray}
We assume that the most rapid variations in the phase field take place across the interfacial region with the direction normal to the interface $\Gamma$. In the limit $\epsilon\ll\delta\ll L$, local to the interface we have \cite{Anderson1996, Lowengrub1998}:
\begin{eqnarray}
&&\nabla \sim \frac{\partial }{\partial \zeta} \bold{\hat{n}}_I,~~~~\nabla c \sim \frac{\partial c}{\partial \zeta} \bold{\hat{n}}_I~~~~{\rm and}~~~~\Delta c \sim \frac{\partial^2 c}{\partial \zeta^2},\label{assump--c2}
\end{eqnarray}
such that
\begin{eqnarray}
\bold{T}=\nabla c \otimes \nabla c\sim\frac{\partial c}{\partial \zeta} \bold{\hat{n}}_I\frac{\partial c}{\partial \zeta} \bold{\hat{n}}_I,~~~~\bold{T}\cdot \bold{\hat{n}}_I\sim\frac{\partial c}{\partial \zeta}\frac{\partial c}{\partial \zeta} \bold{\hat{n}}_I,~~~~{\rm and}~~~~\bold{T}\cdot \bold{\hat{m}}_I\sim 0.\label{simp--stress--tension2}
\end{eqnarray}
Condition (\ref{sidecond}) implies that the fluid velocity term $\rho\bold{v}\otimes(\bold{v}-\bold{v}_{I})\cdot \bold{\hat{n}}_{S}$ is bounded and does not contribute to the integral over the side surface of the pillbox. The terms $-\mu(\nabla \bold{v} +\nabla \bold{v}^{T} )\cdot \bold{\hat{n}}_S $ are bounded and do not contribute to the side integral. We argue that the term $2/3\mu (\nabla \cdot \bold{v})$ is bounded across the interfacial region, and thus does not contribute to the side integral. The pressure $p$ is bounded and does not contribute to the side integral. Further, the non-classical stress term $\bold{T}$ dose not contribute to the integral over the top and bottom surfaces. Eq.(\ref{jump-momen-11}) reduces to
\begin{eqnarray}
&&\int_{\Gamma}\bigg(\big[\rho\bold{v}( \bold{v}-\bold{v}_I)\big]\cdot \bold{\hat{n}}_I+\big[p\bold{I}\big]\cdot \bold{\hat{n}}_I +\big[-\mu(\nabla \bold{v} +\nabla \bold{v}^{T} )\big]\cdot \bold{\hat{n}}_I\bigg){\rm d}S\nonumber\\
&&\hspace{40mm}-\oint_{C}\int_{-\delta}^{\delta}\eta\epsilon\sigma(T)\rho(\frac{\partial c}{\partial \zeta})^2 \bold{\hat{m}}_{I}{\rm d}\zeta{\rm d}l=0,\label{condition---for---mom}
\end{eqnarray}
where the condition (\ref{norm--condition}) is used. Here, for our pillbox argument to make sense, we require that within the pillbox the temperature is continuous and the variations are small over a small distance (of order of the pillbox thickness $\delta$). In the limit $\epsilon\ll \delta \ll L$, the temperature $T$ is approximately uniform along the direction normal to the interface. Note that the similar assumption for the temperature was also suggested by \cite{Jasnow1996}, where a surface tension term with thermocapillary effects was identified from a phase-field model in its sharp-interface limit. Denoting the surface tension by
\begin{eqnarray}
\tilde{\sigma}(T)=\eta\sigma(T)\lim _{\epsilon \to 0}\int_{-\delta}^{\delta}\bigg(\epsilon\rho(\frac{\partial c}{\partial \zeta} )^2\bigg){\rm d} \zeta,\label{surface--term--difinition}
\end{eqnarray}
and substituting into (\ref{condition---for---mom}), we obtain
\begin{eqnarray}
&&\int_{\Gamma}\bigg(\big[\rho\bold{v}( \bold{v}-\bold{v}_I)\big]\cdot \bold{\hat{n}}_I+\big[p\bold{I}\big]\cdot \bold{\hat{n}}_I +\big[-\mu(\nabla \bold{v} +\nabla \bold{v}^{T} )\big]\cdot \bold{\hat{n}}_I\bigg){\rm d}S-\oint_{C}\tilde{\sigma}\bold{\hat{m}}_{I}{\rm d}l=0,\nonumber\\
\label{momentum111}
\end{eqnarray}
where, in the limit $\epsilon\ll \delta \ll L$, we assume that the tangential unit vector $\bold{\hat{m}}_{I}$ is independent of $\zeta$ and thus can be taken out of the line integral. Using the surface divergence theorem \cite{Weatherburn1939} leads to
\begin{eqnarray}
\oint_{C}\tilde{\sigma}\bold{\hat{m}}_{I}{\rm d}l=\int_{\Gamma}\nabla _s \tilde{\sigma} {\rm d} S -\int_{\Gamma}(\nabla _s \cdot \bold{\hat{n}}_I)\tilde{\sigma}\bold{\hat{n}}_I{\rm d}S.\label{surfacedivergence}
\end{eqnarray}
Substituting Eq.(\ref{surfacedivergence}) into Eq.(\ref{momentum111}), we obtain
\begin{eqnarray}
&&\big[\rho\bold{v}( \bold{v}-\bold{v}_I)\big]\cdot \bold{\hat{n}}_I+\big[p\bold{I}\big]\cdot \bold{\hat{n}}_I +\big[-\mu(\nabla \bold{v} +\nabla \bold{v}^{T} )\big]\cdot \bold{\hat{n}}_I=\nabla _s \tilde{\sigma} +\kappa\tilde{\sigma}\bold{\hat{n}}_I.\label{jump--condition--surface--tension--11}
\end{eqnarray}
Here $\nabla _s $ is the surface gradient, $\kappa=-\nabla _s \cdot \bold{\hat{n}}_I $ is the mean curvature of the surface (e.g. \cite{Weatherburn1939}). The first right term is the tangential thermocapillary (Marangoni) force that accounts for the non-uniform surface tension, while the second is the normal surface tension force. Again if we assume that there is no phase change (i.e. no flux) across the interface, Eq.(\ref{jump--condition--surface--tension--11}) reduces to the jump condition that
\begin{eqnarray}
&&\big[p\bold{I}\big]\cdot \bold{\hat{n}}_I +\big[-\mu(\nabla \bold{v} +\nabla \bold{v}^{T} )\big]\cdot \bold{\hat{n}}_I=\nabla _s \tilde{\sigma} +\kappa\tilde{\sigma}\bold{\hat{n}}_I,\label{jump--conditions--surface--tension}
\end{eqnarray}
which is the classical momentum balance jump conditions at the interface for two-phase incompressible fluid with thermocapillary effects.\\
Note that, we can relate the surface tension of our phase-field model $\tilde{\sigma}(T)$ (identified in Eq.(\ref{surface--term--difinition})) to that of the sharp-interface model $\sigma(T)$ (defined in Eq.(\ref{surface--tension--sharpinterface})) by letting
\begin{eqnarray}
\tilde{\sigma}(T)=\eta\sigma(T)\int_{-\delta}^{+\delta}\epsilon\rho(c)(\frac{{\rm d} c}{{\rm d} \zeta})^2{\rm d}\zeta=\sigma(T).\label{excess-sigma4}
\end{eqnarray}
The value of the ratio parameter $\eta$ can then be determined through the following equation
\begin{eqnarray}
\eta=\frac{1}{\int_{-\delta}^{+\delta}\epsilon\rho(c)(\frac{{\rm d} c}{{\rm d} \zeta})^2{\rm d}\zeta}.\label{equ--to--confirm--coef}
\end{eqnarray}
It has been argued in \cite{Chella1996} that in the limit of gently curved interface, and when the motion of the interface is slow, the phase variable $c$ can be approximated by its 1D stationary solution $c_0$ along the direction normal to the interface. For simplicity, we now assume that the local coordinate $\zeta$ coincide with the $y$ direction, and the position of the dividing surface is $y_0=0$.
\begin{figure}
  \centerline{\includegraphics[width=0.5\textwidth]{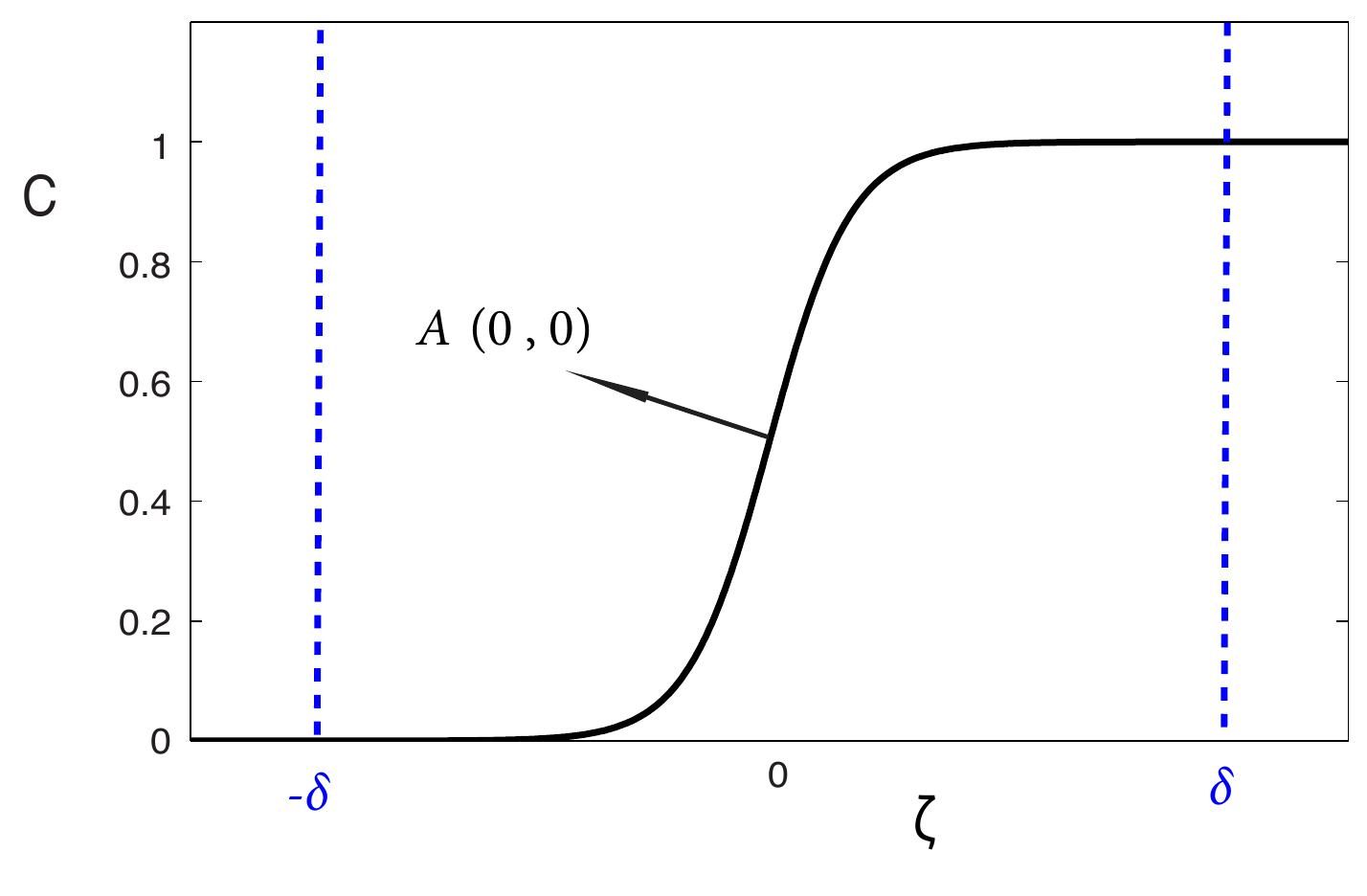}}
  \caption{The stationary solution $c_0$ (black solid line) for the phase field. $A$ is a point on dividing surface $\Gamma$, $\delta$ and $-\delta$ are positions of the top and bottom surfaces of the pillbox (blue dotted line).}\label{convface--2D}
\end{figure}
In 1D case, we have the following stationary solution $c_0$ near the interfacial region,
\begin{eqnarray}
c_0(y)=\frac{1}{2}+\frac{1}{2}{\rm tanh} \big(\frac{y}{2\sqrt{2}\epsilon}\big),~~~~{\rm for}~~y\in[-\delta,\delta],\label{eta--value--1}
\end{eqnarray}
which is shown in Figure \ref{convface--2D}. Here $y=\delta$ and $y=-\delta$ are the positions of the top and bottom surface of the pillbox separately. In the limit $\epsilon\ll\delta\ll L$, we note the conditions
\begin{eqnarray}
c=0~~~~{\rm for}~~~~y=\delta,~~~~{\rm and}~~~~c=1~~~~{\rm for}~~~~y=- \delta.\label{eta--value--2}
\end{eqnarray}
Substituting Eq.(\ref{eta--value--1}) and the variable density (\ref{variable--c}) into (\ref{equ--to--confirm--coef}) we obtain
\begin{eqnarray}
\eta=\frac{2\sqrt{2}(\rho_2-\rho_1)^3}{\rho_1\rho_2\big[\rho_2^2-\rho_1^2-2\rho_1\rho_2   {\rm ln}(\frac{\rho_2}{\rho_1})\big]},
\end{eqnarray}
where the condition (\ref{eta--value--2}) is used. Note that for the density matched case ($\rho_1=\rho_2$), Eq.(\ref{equ--to--confirm--coef}) leads to a simpler expression for $\eta$, which is
\begin{eqnarray}
\eta=6\sqrt{2}.\label{constant--value--eta}
\end{eqnarray}
This agrees with the result obtained in \cite{Rowlinson1982}, \cite{Yue2004} and \cite{Alandthesis}. In $\S6$, we will compute some examples by using our phase-field model for quasi-incompressible fluids (\ref{sys--quasi--mass})-(\ref{sys--quasi--phi2}).
\subsection{Jump condition for energy balance}
To derive the jump condition for energy balance at the interface, we first substitute the terms $E$, $\bold{m}$, $\bold{q}_{E}$ and $\bold{q}^{nc}_{E}$ ((\ref{vol--com--en}), (\ref{energgg1}), (\ref{energgg2}) and (\ref{energgg3})) into the energy balance equation (\ref{der--com--en}). In the integral form, we obtain
\begin{eqnarray}
&&\int_{S}\bigg(\rho \hat{u} (\bold{v}-\bold{v}_{I})+ \rho \frac{1}{2} |\bold{v}|^2 (\bold{v}-\bold{v}_{I})+\big(p\bold{I}-\eta\epsilon\sigma(T)\rho|\nabla c|^2\bold{I}+\eta\epsilon\sigma(T)\rho\bold{T} -\mu(\nabla \bold{v} +\nabla \bold{v}^{T} )\nonumber\\
&&+\frac{2}{3}\mu (\nabla \cdot \bold{v})\bold{I}\big)\cdot \bold{v}-\lambda_{u} (\rho\nabla c\frac{{\rm D}c}{{\rm D}t} )-k\nabla T-m_{C}\mu_{C}\nabla \mu_{C}\bigg)\cdot \bold{n}_{I}{\rm d}S=0,\label{jump--heat1}
\end{eqnarray}
where we have used the identities
\begin{eqnarray}
&&\rho\frac{{\rm D} \hat{u}}{{\rm D} t}=\frac{\partial (\rho\hat{u} )}{\partial t}+\nabla \cdot (\rho\hat{u} \bold{v}),\\
&&\rho\frac{{\rm D} }{{\rm D} t}\frac{1}{2}|\bold{v}|^2=\frac{\partial }{\partial t}(\rho\frac{1}{2}|\bold{v}|^2)+\nabla \cdot (\rho\frac{1}{2}|\bold{v}|^2\bold{v}),
\end{eqnarray}
and the following properties which are similar to (\ref{jump--mass51121}) and (\ref{jump--mass51122}),
\begin{eqnarray}
&&\int_{V}\frac{\partial (\rho\hat{u})}{\partial t}{\rm d}V\sim-\int_{S}\rho\hat{u}\bold{v}_{I}\cdot \bold{\hat{n}}{\rm d}S,\label{jump--mass51123}\\
&&\int_{V}\frac{\partial}{\partial t}  (\rho\frac{1}{2}|\bold{v}|^2) {\rm d }V\sim-\int_{S} \rho\frac{1}{2}|\bold{v}|^2\bold{v}_{I}\cdot \bold{\hat{n}}{\rm d}S.\label{time2}
\end{eqnarray}
We then break up the above integral (\ref{jump--heat1}) into pieces for the top, bottom and sides of the pillbox to obtain
\begin{eqnarray}
&&\int_{S_{top}}\bigg(\rho \hat{u} (\bold{v}-\bold{v}_{I}) + \rho \frac{1}{2} |\bold{v}|^2 (\bold{v}-\bold{v}_{I}) +\big(p\bold{I}-\eta\epsilon\sigma(T)\rho|\nabla c|^2\bold{I}+\eta\epsilon\sigma(T)\rho\bold{T} \nonumber\\
&&-\mu(\nabla \bold{v} +\nabla \bold{v}^{T} )+\frac{2}{3}\mu (\nabla \cdot \bold{v})\bold{I}\big)\cdot \bold{v}-\lambda_{u} (\rho\nabla c\frac{{\rm D}c}{{\rm D}t} )-  k\nabla T-m_{C}\mu_{C}\nabla \mu_{C}\bigg)\cdot \bold{\hat{n}}_{T}{\rm d}S\nonumber\\
&&+\int_{S_{bot}}\bigg(\rho \hat{u} (\bold{v}-\bold{v}_{I})+\rho \frac{1}{2} |\bold{v}|^2 (\bold{v}-\bold{v}_{I})  +\big(p\bold{I}-\eta\epsilon\sigma(T)\rho|\nabla c|^2\bold{I}+\eta\epsilon\sigma(T)\rho\bold{T} \nonumber\\
&&-\mu(\nabla \bold{v} +\nabla \bold{v}^{T} )+\frac{2}{3}\mu (\nabla \cdot \bold{v})\bold{I}\big)\cdot \bold{v}-\lambda_{u} (\rho\nabla c\frac{{\rm D}c}{{\rm D}t} )- k\nabla T-m_{C}\mu_{C}\nabla \mu_{C}\bigg)\cdot \bold{\hat{n}}_{B}{\rm d}S\nonumber\\
&&+\oint_{C}\int_{-\delta}^{\delta}\bigg(\rho \hat{u} (\bold{v}-\bold{v}_{I})+\rho \frac{1}{2} |\bold{v}|^2 (\bold{v}-\bold{v}_{I}) +\big(p\bold{I}-\eta\epsilon\sigma(T)\rho|\nabla c|^2\bold{I}+\eta\epsilon\sigma(T)\rho\bold{T} \nonumber\\
&&-\mu(\nabla \bold{v} +\nabla \bold{v}^{T} )+\frac{2}{3}\mu (\nabla \cdot \bold{v})\bold{I}\big)\cdot \bold{v}-\lambda_{u} (\rho\nabla c\frac{{\rm D}c}{{\rm D}t} )-k\nabla T-m_{C}\mu_{C}\nabla \mu_{C}\bigg)\cdot \bold{\hat{n}}_{S}{\rm d}\zeta{\rm d}l=0,\nonumber\\
&&\label{jump--heat22}
\end{eqnarray}
where we assume that the heat capacity $c_{hc}$ is a constant. In the limit $\epsilon\ll \delta\ll L$, the non-classical terms of the internal energy $\hat{u}$ (Eq.(\ref{specified--internal--energy1})), $\bold{T}$, $\lambda_{u}(\rho\nabla c \frac{{\rm D}c}{{\rm D}t})$ and $m_{C}\mu_{C}\nabla\mu_{C}$ do not contribute to the top and the bottom surface integrals. The non-classical term $m_{C}\mu_{C}\nabla \mu_{C}$ and the energy term $k\nabla T$ are bounded in the tangential direction and do not contribute to the side integral. Eq.(\ref{jump--heat22}) then reduces to
\begin{eqnarray}
&&\int_{\Gamma}\bigg(\big[\rho c_{hc}T (\bold{v}-\bold{v}_{I})\big]+\big[\rho\frac{1}{2} |\bold{v}|^2 (\bold{v}-\bold{v}_{I})\big]+ \big[p\bold{I}\cdot \bold{v}\big] -\big[\mu(\nabla \bold{v} +\nabla \bold{v}^{T} )\cdot \bold{v}\big]\nonumber\\
&&-  \big[k\nabla T\big]\bigg)\cdot \bold{\hat{n}}_{I}{\rm d}S-\oint_{C}\bigg(\int_{-\delta}^{\delta} \eta\epsilon\sigma(T)\rho(\frac{{\rm d}c}{{\rm d}\zeta})^2{\rm d}\zeta\bigg)\bold{v}_{I}\cdot \bold{\hat{m}}_{I}{\rm d}l=0,\label{dfadf}
\end{eqnarray}
where in the last term of Eq.(\ref{dfadf}), we argue that the interface velocity $\bold{v}_{I}$ is independent of the local coordinate $\zeta$ and thus can be taken out of the integral in the normal direction. By using Eq.(\ref{surface--term--difinition}) and the surface divergence theorem, we obtain
\begin{eqnarray}
&&\big[k\nabla T\big]\cdot \bold{\hat{n}}_{I}=\bigg(\big[\rho c_{hc}T (\bold{v}-\bold{v}_{I})\big]+\big[ \rho \frac{1}{2} |\bold{v}|^2 (\bold{v}-\bold{v}_{I})\big]+ \big[p\bold{I}\cdot \bold{v}\big] \nonumber\\
&&- \big[\mu(\nabla \bold{v} +\nabla \bold{v}^{T} )\cdot \bold{v}\big] \bigg)\cdot \bold{\hat{n}}_{I} -\nabla _s\cdot (\tilde{\sigma}\bold{v}_{I} ) - \kappa\tilde{\sigma}\bold{\hat{n}}_I \cdot \bold{v}_{I},
\label{jump--heat711}
\end{eqnarray}
where the energy spent by the interface deformation and the effects of the interface curvature are taken into account in our jump condition for energy balance at the interface. Eq.(\ref{jump--heat711}) agrees with the result obtained in \cite{Andrea2011}, where the energy balance condition at the interface is derived by using a pillbox for sharp interface model. Again if we assume that there is no phase change across the interface, Eq.(\ref{jump--heat711}) then reduces to
\begin{eqnarray}
&&\big[k\nabla T\big]\cdot \bold{\hat{n}}_{I}=\bigg( \big[p\bold{I}\cdot \bold{v}\big] - \big[\mu(\nabla \bold{v} +\nabla \bold{v}^{T} )\cdot \bold{v}\big] \bigg)\cdot \bold{\hat{n}}_{I} -\nabla _s\cdot (\tilde{\sigma}\bold{v}_{I} ) -\kappa\tilde{\sigma}\bold{\hat{n}}_I \cdot \bold{v}_{I}.\nonumber\\
\label{jump--heat711222}
\end{eqnarray}
If we further ignore the energy spent by the interface deformation and the effects of interface curvature, we can obtain the classical jump condition for the energy equation,
\begin{eqnarray}
\big[k\nabla T\big]\cdot \bold{\hat{n}}_I=0,\label{jump--heat71}
\end{eqnarray}
which is widely used for the computations of sharp-interface model (e.g. \cite{Tavener2002}).
\Section{Computational methods and results}
In this section, we investigate numerically our phase-field model through three examples. One is the thermocapillary convection in a micro-channel with two-layer superimposed fluid, and the second (third) one is the thermocapillary migration of a drop with zero (finite, respectively) Marangoni number. All examples will be computed by using continuous finite element methods. The numerical results of the first and second examples will be compared to the existing analytical solutions and numerical results.
\subsection{Simplified model and the weak form}
For the sake of simplicity, we assume that the densities of the two fluids are matched. The system equations (\ref{sys--quasi--mass})-(\ref{sys--quasi--phi2}) can then be simplified in the form
\begin{align}
\nabla \cdot \bold{v}&=0,\label{211sp--sys--quasi--mass}\\
\frac{{\rm D} \bold{v}}{{\rm D} t}&=-\nabla p+\nabla \big[\lambda_f(T)|\nabla c|^2\big]-\nabla \cdot \big[\lambda_f(T)(\nabla c\otimes \nabla c)\big] +\nabla \cdot (\mu \nabla \bold{v}),\label{211sp--sys--quasi--mom}\\
\frac{{\rm D} \hat{u}}{{\rm D} t}&=\lambda_{u}\nabla \cdot (\nabla c\frac{{\rm D}c}{{\rm D}t} )+\big(- p\bold{I}+\lambda_{f}(T)|\nabla c|^2\bold{I}-\lambda_{f}(T)(\nabla c\otimes \nabla c)+ \mu\nabla\bold{v} \big): \nabla \bold{v}\nonumber\\
&+\nabla \cdot  (k\nabla T+m_{C}\mu_{C}\nabla \mu _{C}),\label{211sp--sys--quasi--en}\\
\frac{{\rm D}   c}{{\rm D} t}&=m_C\Delta \mu_{C},\label{211sp--sys--quasi--phi1}\\
\mu_{C} &=\gamma_f(T)\frac{{\rm d} h(c)}{{\rm d} c} -\lambda_{f}(T)\Delta c,\label{211sp--sys--quasi--phi2}
\end{align}
where the variable thermal conductivity (\ref{variable--thermaldiffusicity}) is employed. Here we employed the energy balance equation (\ref{der--com--en}) instead of (\ref{sys--quasi--en}). The reason is that in the weak formulation of Eq.(\ref{sys--quasi--en}), the second order derivative is involved implying that more reductive $C^1$ finite elements are needed for the conformity. However in the weak formulation of (\ref{211sp--sys--quasi--en}) ( (\ref{w--sys--quasi--en}) below) we find that only first order derivatives of $c$ are involved, so that the $C^0$ finite element method may be used for our computations. The benefits of using $C^0$ elements are obvious, that the method can have more choices of elements and many existing codes can be incorporated to reduce various complications. Note that Eqs.(\ref{211sp--sys--quasi--mass})-(\ref{211sp--sys--quasi--en}) of the system will be computed for the example of thermocapillary convection, Eqs.(\ref{211sp--sys--quasi--mass})-(\ref{211sp--sys--quasi--phi2}) will be computed for the example of thermocapillary migration with zero Marangoni number, and the non-dimensional system equations (\ref{non--sys--quasi--mass})-(\ref{non--sys--quasi--en}) will be computed for the example of thermocapillary migration with finite Marangoni number. For simplicity, we only present the numerical scheme for dimensional system equations (\ref{211sp--sys--quasi--mass})-(\ref{211sp--sys--quasi--phi2}). The numerical method for the non-dimensional system (\ref{non--sys--quasi--mass})-(\ref{non--sys--quasi--en}) can be obtained correspondingly. By multiplying the system (\ref{211sp--sys--quasi--mass})-(\ref{211sp--sys--quasi--phi2}) with the test functions $q$, $\bold{u}$, $\chi$, $\phi$ and $\psi$ respectively and using integration by parts, the weak form can be derived straightforwardly (where $\bold{v}$, $p$, $\hat{u}$, $c$, $\mu$ and test functions $\bold{u}$, $q$, $\chi$, $\phi$ and $\psi$ are in appropriate spaces),
\begin{eqnarray}
&&\int_{\Omega}\bigg(\nabla \cdot \bold{v}q\bigg){\rm d}\bold{x}=0,\label{w--sys--quasi--mass}\\
&&\int_{\Omega}\bigg(\bold{v}_t \cdot \bold{u}+  (\bold{v}\cdot \nabla )\bold {v}\cdot \bold{u}-p\nabla \cdot \bold{u}+ \lambda_f(T)(\nabla c\cdot \nabla c)\nabla \cdot \bold{u}-  \lambda_f(T)(\nabla c\otimes \nabla c) : \nabla \bold{u}\nonumber\\
&&+ \mu \nabla \bold{v}:\nabla \bold{u}\bigg){\rm d}\bold{x}=0,\label{w--sys--quasi--mom}
\end{eqnarray}
\begin{eqnarray}
&&\int_{\Omega}\bigg( \hat{u}_t\chi+(\bold{v}\cdot \nabla)\hat{u}\chi + \lambda_{u}\frac{{\rm D}c}{{\rm D}t} \nabla c\cdot \nabla \chi   -\lambda_f (T)(\nabla c\cdot \nabla c) (\nabla\cdot \bold{v})\chi-\mu \nabla \bold{v} : \nabla \bold{v}\chi \nonumber\\
&&+\lambda_f (T)(\nabla c\otimes \nabla c) : \nabla \bold{v}\chi+p\bold{I}:\nabla \bold{v}\chi+ k\nabla T\cdot \nabla \chi +m_{C}\mu_{C}\nabla \mu _{C}\cdot \nabla \chi\bigg){\rm d}\bold{x}=0,\label{w--sys--quasi--en}\\
&&\int_{\Omega}\bigg( c_t\phi+(\bold{v}\cdot \nabla)c\phi +m_C\nabla \mu_{C}\cdot \nabla \phi\bigg){\rm d}\bold{x}=0,\label{w--sys--quasi--phi1}\\
&&\int_{\Omega}\bigg(\mu_{C}\psi -\gamma_f(T)\frac{{\rm d} h(c)}{{\rm d} c}\psi -\nabla \lambda_{f}(T) \cdot \nabla  c\psi- \lambda_{f}(T) \nabla  c\cdot \nabla \psi\bigg){\rm d}\bold{x}=0.\label{w--sys--quasi--phi2}
\end{eqnarray}
\subsection{Temporal schemes and implement issue}
The solution of the weak form (\ref{w--sys--quasi--mass})-(\ref{w--sys--quasi--phi2}) is approximated by a finite difference scheme in time and a conformal $C^0$ finite element method in space. To ensure the stability of our numerical method, we adopt the fully implicit backward Euler scheme to compute the problem.\\
We let $\Delta t >0$ represent a time step size, and ($\bold{v}_{h}^{n}, p_{h}^{n}, \hat{u}_{h}^{n}, c_{h}^{n}, {\mu_{C}}_{h}^{n}$) (in a finite dimensional space given by a finite element discretization of the computational domain $\Omega$) is an approximation of $(\bold{v}, p, \hat{u}, c, \mu)$ at time $t^{n}=n\Delta t$, where ${\bold{v}}^n_{h}=\bold{v}(n\Delta t)$, $p^n_{h}=p(n\Delta t)$, ${\hat{u}}^n_{h}=\hat{u}(n\Delta t)$, $c^n_{h}=c(n\Delta t)$ and ${\mu_{C}}^n_{h}=\mu_{C}(n\Delta t)$. Then the approximation at time $t^{n+1}$ is denoted as ($\bold{v}_{h}^{n+1}, p_{h}^{n+1}, \hat{u}_{h}^{n+1}, c_{h}^{n+1}, {\mu_{C}}_{h}^{n+1}$) and computed by the following finite element scheme
\begin{eqnarray}
&&\int_{\Omega}\bigg(\nabla \cdot \bold{v}_{h}^{n+1}q+\delta p^{n+1}_{h}q\bigg){\rm d}{\bold x}=0,\label{tsw--sys--quasi--mass}\\
&&\int_{\Omega}\bigg(\bold{v}^{n+1}_{\bar{t}} \cdot \bold{u}+  (\bold{v}_{h}^{n+1}\cdot \nabla )\bold {v}_{h}^{n+1}\cdot \bold{u}-p_{h}^{n+1}\nabla \cdot \bold{u}+ \lambda_f(T_{h}^{n+1})(\nabla c_{h}^{n+1}\cdot\nabla c_{h}^{n+1})  \nabla\cdot \bold{u}\nonumber\\
&&-  \lambda_f(T_{h}^{n+1})(\nabla c_{h}^{n+1}\otimes \nabla c_{h}^{n+1}) : \nabla \bold{u}+ \mu \nabla \bold{v}_{h}^{n+1}:\nabla \bold{u}\bigg){\rm d}{\bold x}=0,\label{tsw--sys--quasi--mom}\\
&&\int_{\Omega}\bigg( \hat{u}^{n+1}_{\bar{t}}\chi+(\bold{v}_{h}^{n+1}\cdot \nabla)\hat{u}_{h}^{n+1}\chi + \lambda_{u} (c^{n+1}_{\bar{t}}+(\bold{v}_{h}^{n+1}\cdot \nabla)c_{h}^{n+1}) \nabla c_{h}^{n+1}\cdot \nabla \chi    \nonumber\\
&&-\lambda_f (T_{h}^{n+1})(\nabla c_{h}^{n+1}\cdot \nabla c_{h}^{n+1}) (\nabla \cdot \bold{v}_{h}^{n+1})\chi+\lambda_f (T_{h}^{n+1})(\nabla c_{h}^{n+1}\otimes \nabla c_{h}^{n+1}) : \nabla \bold{v}_{h}^{n+1}\chi \nonumber\\
&&-\mu \nabla \bold{v}_{h}^{n+1} : \nabla \bold{v}_{h}^{n+1}\chi+ k\nabla T_{h}^{n+1}\cdot \nabla \chi +m_{C}{\mu_{C}}_{h}^{n+1}\nabla {\mu_{C}}_{h}^{n+1}\cdot \nabla \chi\bigg){\rm d}\bold{x}=0,\label{w--sys--quasi--en1212}\\
&&\int_{\Omega}\bigg( c^{n+1}_{\bar{t}}\phi+(\bold{v}_{h}^{n+1}\cdot \nabla)c_{h}^{n+1}\phi  +m_C\nabla {\mu_{C}}_{h}^{n+1}\cdot \nabla \phi\bigg){\rm d}{\bold x}=0,\label{tsw--sys--quasi--phi1}\\
&&\int_{\Omega}\bigg({\mu_{ C}}_{h}^{n+1}\psi -\gamma_f(T_{h}^{n+1}) h'(c_{h}^{n+1})\psi -\nabla \lambda_{f}(T_{h}^{n+1}) \cdot \nabla  c_{h}^{n+1}\psi\nonumber\\
&&- \lambda_{f}(T_{h}^{n+1}) \nabla  c_{h}^{n+1}\cdot \nabla \psi\bigg){\rm d}{\bold x}=0,\label{tsw--sys--quasi--phi2}
\end{eqnarray}
where $\bold{v}^{n+1}_{\bar{t}}={(\bold{v}^{n+1}_{h}-\bold{v}^{n}_{h})}/{\Delta t}$, $\hat{u}^{n+1}_{\bar{t}}={(\hat{u}^{n+1}_{h}-\hat{u}^{n}_{h})}/{\Delta t}$ and $c^{n+1}_{\bar{t}}={(c^{n+1}_{h}-c^{n}_{h})}/{\Delta t}$. Note that the divergence free equation needs to be treated carefully in incompressible flow computations. Here we rewrite Eq.(\ref{tsw--sys--quasi--mass}) in the penalty formulation, where $\delta$ is a relatively small parameter and is set to be $\delta=10^{-6}$ for all the computations. Note that for every time step, $T^{n+1}$ can be obtained by using Eq.(\ref{specified--internal--energy1}), such that
\begin{eqnarray}
c_{hc}T_{h}^{n+1}={\hat{u}}_{h}^{n+1}-\gamma _u h(c_{h}^{n+1})-\lambda_{u}\frac{1}{2}\nabla c_{h}^{n+1}\cdot \nabla c_{h}^{n+1}.
\end{eqnarray}
Since the scheme is nonlinearly implicit we need to do the linearization and then solve a linear system iteratively at each time step. We follow the numerical methods designed by Hua $et$ $al.$ \cite{Hua2011}, where the linear system is symmetric and does not depend on time. Therefore, we only need to do the Cholesky factorization for the symmetric linear system at the initial time step. After the initial time we do not need to factorize the linear system again since the coefficient matrix is independent of time.\\
For a phase-field model, it is sufficient to finely resolve only the interfacial region, and a fixed grid meshing represents a waste of computational resources. Therefore, efficient adapting mesh that resolves the thin interfacial region is desirable. For the examples of the thermocapillary convection, we design a mesh that has relatively high-resolution grids near the flat interface. For the example of the thermocapillary migration, since the interface moves as the drop rises, an adaptive mesh is designed, in which there is a smaller frame that moves with the drop. Within the frame, the resolution of grids is much higher than those outside the moving frame, so that the moving interface of the drop can be resolved purposely. Here, only the meshes for the example of thermocapillary migration are shown.
\subsection{Thermocapillary convection in a two-layer fluid system}
\begin{figure}
  \centerline{\includegraphics[width=0.6\textwidth]{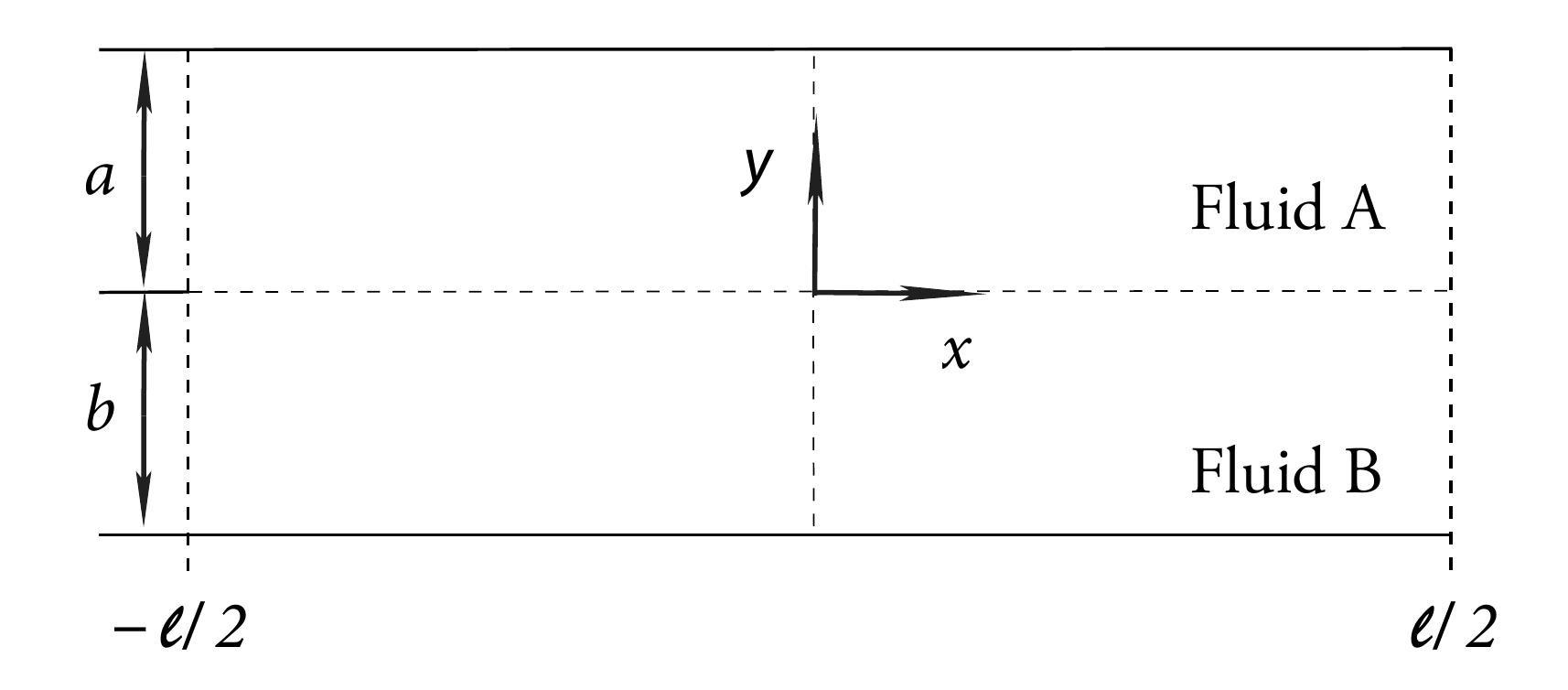}}
                \caption{The schematic diagram showing two immiscible fluids in a microchannel. The temperatures of the lower and upper plates are $T^{b}(x, -b) = T_h + T_0 {\rm cos} (kx)$ and $T^{a}(x, a) = T_c$, respectively, where $T_h > T_c > T_0$ and $k=2\pi/l$ is the wave number, and $a$ and $b$ are the heights of the fluid A and B respectively.}\label{convection11--box}
\end{figure}

We now investigate the thermocapillary convection in a heated micro-channel with two-layer superimposed fluids with a planar interface \cite{Pendse2010}. Considering two-layer fluids (Figure {\ref{convection11--box}}), where the heights of the fluid A (upper) and fluid B (lower) are $a$ and $b$, respectively, and the fluids are of infinite extension in the horizontal direction. The physical properties of the fluids are their densities, viscosities and heat conductivities. The temperature variations in the present study are considered to be small enough so that the thermophysical properties of each fluid are assumed to remain constant, with the exception of surface tension. The temperature of the lower and upper plates are 
\begin{eqnarray}
T^{b}(x,-b)=T_h+T_0~{\rm cos}(\omega x)~~~~{\rm and}~~~~T^a(x,a)=T_c\label{temp--boundary--conditions}
\end{eqnarray}
respectively, where $T_h>T_c>T_0>0$, and $\omega=2\pi /l$ is a wave number with $l$ being the channel length. The above temperature boundary conditions establish a temperature field that is periodic in the horizontal direction with a period of $l$. Therefore, it is only sufficient to focus on the solution in one period, i.e.,$-l/2<x<l/2$.
In the limit of zero Marangoni number and small Reynolds number, it is possible to ignore the convective transport of momentum and energy. In addition, we assume that the interface is to remain flat. By solving the simplified sharp-interface governing equations with the corresponding jump boundary conditions at the interface, \cite{Pendse2010} obtained the analytical solutions for temperature field $\bar{T}(x,y)$ and stream-function $\bar{\psi}(x,y)$, where for the upper fluid
\begin{eqnarray}
&&\bar{T}^{A}(x,y)=\frac{(T_c-T_h)y+\tilde{k}T_cb+T_ha}{a+\tilde{k}b}+T_0f(\alpha,\beta,\tilde{k})~{\rm sinh}(\alpha-\omega y)~{\rm cos}(\omega x),\\
&&\bar{\psi}^{A}(x,y)=\frac{U_{max}}{\omega}\frac{1}{{\rm sinh}^2(\alpha)-\alpha^2}\big\{\omega y~{\rm sinh}^2(\alpha){\rm cosh}(\omega y)\nonumber\\
&&\hspace{35mm}-\frac{1}{2}\big[ 2\alpha^2+\omega y\big({\rm sinh}(2\alpha)-2\alpha\big)\big]{\rm sinh}(\omega y)\big\}{\rm sin}(\omega x),
\end{eqnarray}
and for the lower fluid
\begin{eqnarray}
&&\bar{T}^{B}(x,y)=\frac{\tilde{k}(T_c-T_h)y+\tilde{k}T_cb+T_ha}{a+\tilde{k}b}+T_0f(\alpha,\beta,\tilde{k})~\big[{\rm sinh}(\alpha){\rm cosh}(\omega y)\nonumber\\
&&\hspace{35mm}-\tilde{k}{\rm sinh}(\omega y) {\rm cosh}(\alpha)\big]~{\rm cos}(\omega x),\\
&&\bar{\psi}^{B}(x,y)=\frac{U_{max}}{\omega}\frac{1}{{\rm sinh}^2(\beta)-\beta^2}\big\{\omega y~{\rm sinh}^2(\beta){\rm cosh}(\omega y)\nonumber\\
&&\hspace{35mm}-\frac{1}{2}\big[ 2\beta^2-\omega y\big({\rm sinh}(2\beta)-2\beta\big)\big]{\rm sinh}(\omega y)\big\}{\rm sin}(\omega x).
\end{eqnarray}
In the above equations the unknowns are defined by $\tilde{k}={k_A}/{k_B}$, $\alpha=a\omega$, $\beta=b\omega$, $f(\alpha,\beta,\tilde{k})={1}/({\tilde{k}{\rm sinh}(\beta){\rm cosh}{\alpha}+{\rm sinh}(\alpha){\rm cosh}{\beta}})$, $g(\alpha,\beta,\tilde{k})={\rm sinh}(\alpha)f(\alpha,\beta,\tilde{k})$
and
\begin{eqnarray}
&&\hspace{30mm} U_{max}=-\bigg(\frac{T_0\sigma_T}{\mu_B}\bigg)g(\alpha,\beta,\tilde{k})h(\alpha,\beta,\tilde{\mu}),\nonumber\\
&&h(\alpha,\beta,\tilde{\mu})=\frac{\big({\rm sinh}^2(\alpha)-\alpha^2\big)\big({\rm sinh}^2(\beta)-\beta^2\big)}{\tilde{k}\big({\rm sinh}^2(\beta)-\beta^2\big)\big({\rm sinh}(2\alpha)-2\alpha\big)+\big({\rm sinh}^2(\alpha)-\alpha^2\big)\big({\rm sinh}(2\beta)-2\beta\big)}.\nonumber
\end{eqnarray}
Based on their work, the simulations for our phase-field model are carried out in a 2D domain $[-l/2,l/2]\times [-b,a]$ with $l=1.6\times10^{-4}$, and $a=b=4\times10^{-5}$. As the interface between the two fluids is assumed to be flat and rigid, the initial conditions for the phase variable are only depending on $y$, and can be given in the form
\begin{eqnarray}
c(y)=\frac{1}{2}+\frac{1}{2}{\rm tanh} \big(\frac{y}{2\sqrt{2}\epsilon}\big),~~~~{\rm for}~~y\in(-b,a).\label{incon--c}
\end{eqnarray}
The periodic boundary conditions are applied on the left and right sides of the domain. On the top and bottom walls, the no-slip boundary conditions are imposed such that 
\begin{eqnarray}
\bold{v}=0~~~~{\rm for}~~~~y=a,-b.
\end{eqnarray}
Eq.(\ref{temp--boundary--conditions}) are used as the boundary conditions for temperature with $T_{h}=20$, $T_{c}=10$ and $T_0=4$. We let the ratio parameter $\eta=6\sqrt{2}$ (Eq. (\ref{constant--value--eta})). Moreover, the fluid properties are shown in Table \ref{tab:kd11}.
\begin{table}
\begin{center}
\begin{tabular}{ll}
$\mu_A=\mu_B=0.2,$  & \hspace{-20mm}$k_B=0.2,$\hspace{30.5mm}$\sigma_0=2.5\times10^{-1},$\\
$\tilde{k}=k_A/k_B~~({\rm thermal~conductivity~ratio}),$ \hspace{5mm} & $\sigma_T=-5\times10^{-3}~~({\rm at}~~T_{ref}=T_c)$,\\
\end{tabular}
  \caption{The physical properties of two fluids for example of thermocapillary convection (A and B stand for the fluid A and B separately).}
  \label{tab:kd11}
\end{center}
\end{table}
To show the influences of the thermal conductivity ratio on the stream-function and temperature fields, the simulations are carried out for two cases with different values of $\tilde{k}$, where $\tilde{k}=0.1$ for case 1, and $\tilde{k}=0.5$ for case 2. Here the variable thermal conductivity $k(c)$ (Eq. (\ref{variable--thermaldiffusicity})) is employed, where we fix $k_B(=0.2)$, and change the value of $k_A$ for the two cases.  
\begin{figure}
  \centerline{\includegraphics[width=0.95\textwidth]{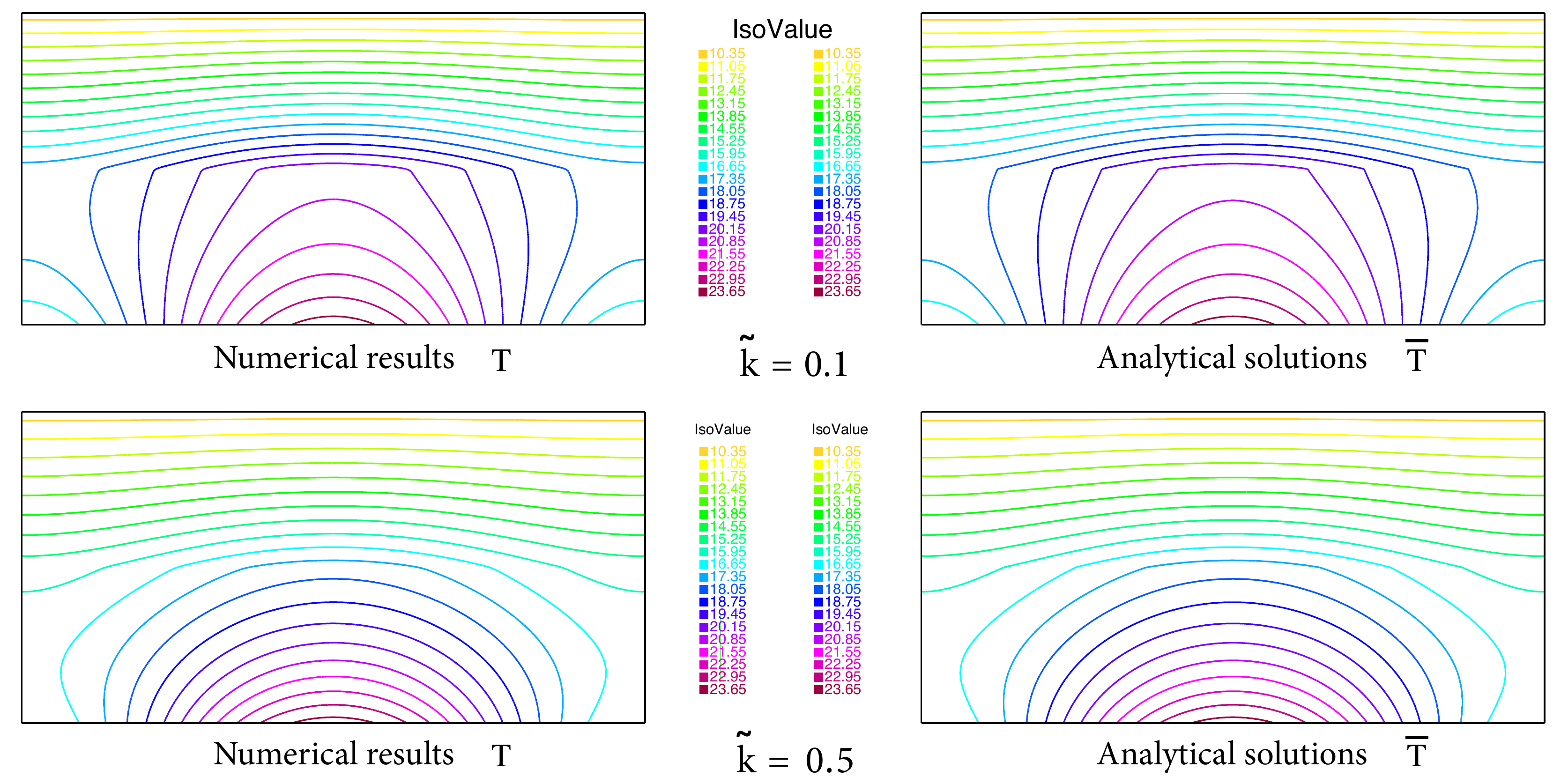}}
                \caption{Isotherms of the numerical results and analytical solutions for the example of thermocapillary convection in a two-layer fluid system with the different thermal diffusivity ratios, $\tilde{k}=0.1$ and $\tilde{k}=0.5$.}\label{thermal--temperature--field1}    
\end{figure}
\begin{figure}
  \centerline{\includegraphics[width=0.95\textwidth]{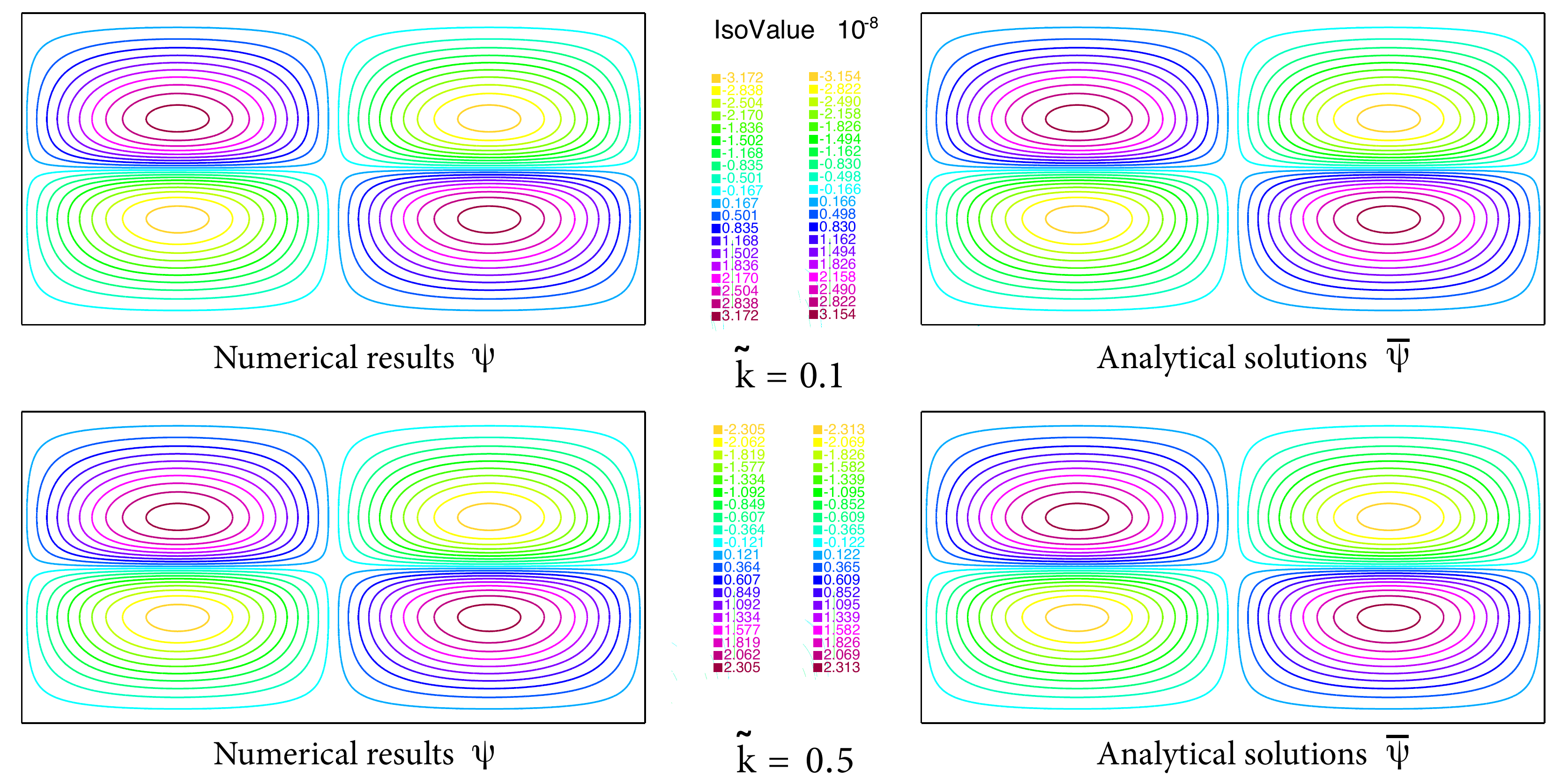}}
                \caption{Streamlines of the numerical results and analytical solutions for the example of thermocapillary convection in two-layer fluid system with different thermal diffusivity ratios, $\tilde{k}=0.1$ and $\tilde{k}=0.5$. Positive (negative) values of the stream-function indicate the clockwise (the counterclockwise) circulation.}\label{thermal--streamfunction--field1}          
\end{figure}
\begin{table}
\begin{center}
\begin{tabular}{lclllll}
~&~&$\epsilon=0.02$ & $\epsilon=0.01$\hspace{5mm}&$\epsilon=0.005$\hspace{5mm}&$\epsilon=0.002$\hspace{5mm}&$\epsilon=0.001$\hspace{5mm}\\
k=0.1&$\frac{||T-\bar{T}||_{L^{2}}}{||\bar{T}||_{L^{2}}}$& $5.445\times 10^{-3}$&$2.503\times 10^{-3}$ \hspace{5mm}&$1.189\times 10^{-3}$\hspace{5mm}&$4.551\times 10^{-4}$\hspace{5mm}&$2.200\times 10^{-4}$\\
~&$\frac{||\psi-\bar{\psi}||_{L^{2}}}{||\bar{\psi}||_{L^{2}}}$&$4.309\times 10^{-2}$ &$2.668\times 10^{-2}$ \hspace{5mm}&$1.614\times 10^{-2}$\hspace{5mm}&$6.94\times 10^{-3}$\hspace{5mm}&$6.44\times 10^{-3}$\\
~&~&~&~&~&~\\
k=0.5&$\frac{||T-\bar{T}||_{L^{2}}}{||\bar{T}||_{L^{2}}}$&$1.585\times 10^{-3}$ & $5.748\times 10^{-4}$\hspace{5mm}&$2.098\times 10^{-4}$\hspace{5mm}&$5.167\times 10^{-5}$\hspace{5mm}&$1.815\times 10^{-5}$\\
~&$\frac{||\psi-\bar{\psi}||_{L^{2}}}{||\bar{\psi}||_{L^{2}}}$&$6.796\times 10^{-2}$ & $2.208\times 10^{-2}$\hspace{5mm}&$8.682\times 10^{-3}$\hspace{5mm}&$3.688\times 10^{-3}$\hspace{5mm&$7.318\times 10^{-4}$}\\
\end{tabular}
  \caption{$L^{2}$ norm of the relative differences between the numerical results and the analytical solutions for $\S 6.3$.}
  \label{tab:kd11aaa}
\end{center}
\end{table}
The contours of temperature fields and stream function for two cases at $\epsilon=0.002$ are shown in Figure \ref{thermal--temperature--field1} and \ref{thermal--streamfunction--field1} respectively. It can be seen that our numerical results are in good agreement with the analytical solutions. In order to show that our phase-field model approaches to the sharp-interface model as the thickness of diffuse interface goes to zero, the computations are carried out by using five different values of $\epsilon$($=0.02, 0.01, 0.005, 0.002, 0.001$). The $L^{2}$ norm of the relative differences between the numerical results and analytical solutions are shown in Table 2. We can observe that as the value of $\epsilon$ decreases, the $L^{2}$ norm of the relative differences decreases for both temperature field and stream functions. We also note that there are slightly differences between our numerical results and the analytical predictions. The reason is two-fold. For one, and most importantly, the thickness of the interface of our model is finite, and the thermal diffusivity changes across it. The second reason is that the viscous heating term is considered in our energy balance equation (\ref{211sp--sys--quasi--en}). As can be observed from the isotherms in Figure {\ref{thermal--temperature--field1}}, the cosine like boundary condition for temperature leads to the non-uniform distributions of the temperature along the interface. This results in a shear force along the interface that is from the centre to both sides of the domain. The fluids are set to motion by this shear force and move from the middle toward both sides of the domain. It is then replaced by the fluid flowing downwards (upward) from the top (bottom) boundary. Also as the domain is periodic in the horizontal, the velocities of fluid that moves towards both sides decrease and the fluids are forced to move upward (downward) to the top (bottom) of the domain. This mechanism results in the formation of the circulation patterns that can be observed in the stream function fields (Figure \ref{thermal--streamfunction--field1}), where the fluid flow consists of four counter-rotating circulation that divide the domain into four parts. Moreover, in the context of the thermal conductivity ratio, we find that the decrease of $\tilde{k}$ leads to a more non-uniform distribution of temperature along the interface, and thus strengthens the thermocapillary convection. This result agrees with the recent result obtained by Liu $et$ $al.$ \cite{Liu2014}, where the same thermocapillary convection in a two-layer fluid system was investigated numerically by using a lattice Boltzmann phase-field method.
\subsection{Thermocapillary migration in the limit of zero Marangoni number}
\begin{figure}
                \centering
                \includegraphics[width=0.6\textwidth]{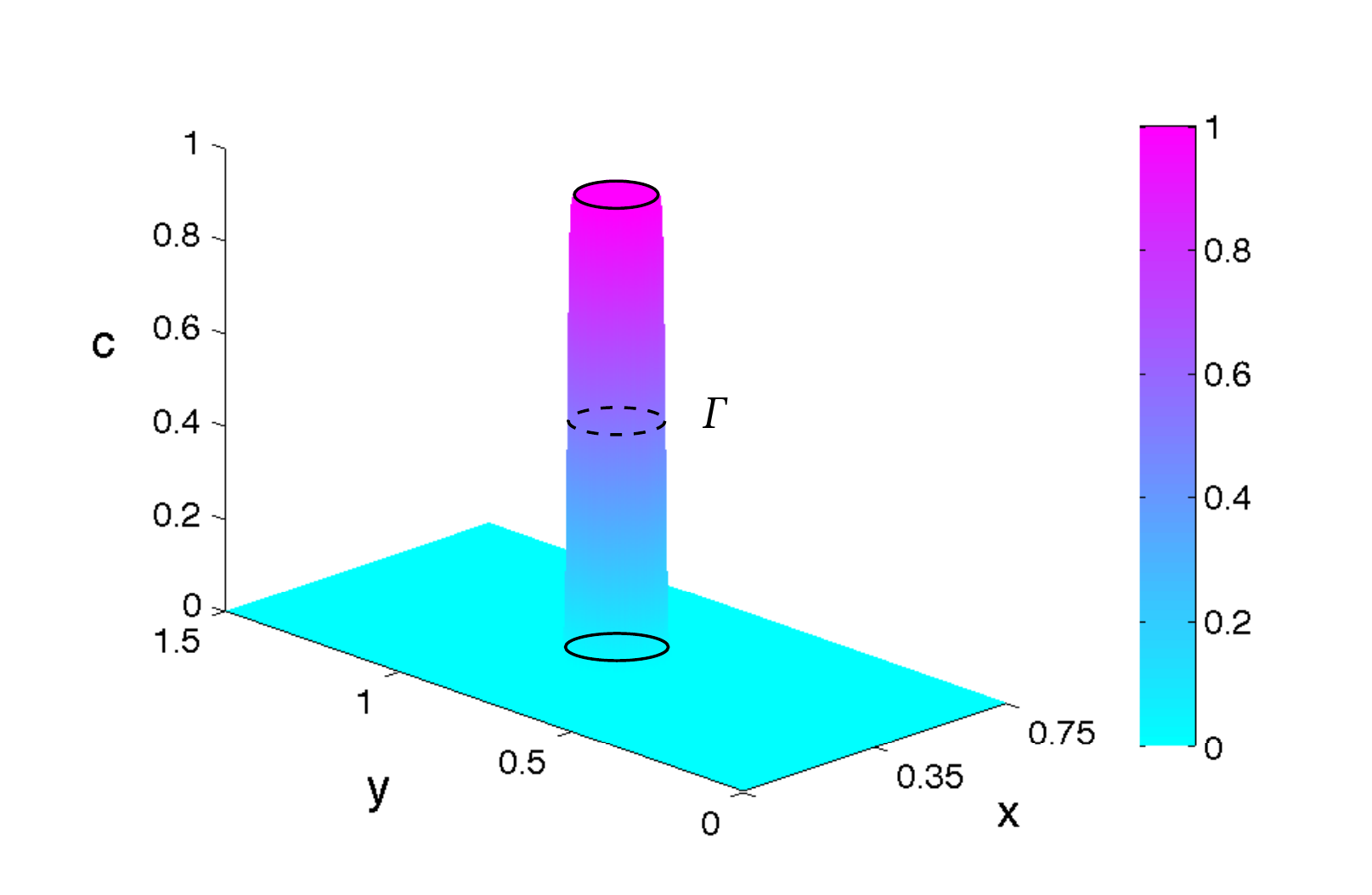}
                \caption{Initial condition of the phase variable $c$ for the example of the thermocapillary migration of a drop. Dotted line stands for the dividing surface.}\label{initial--condition--for--concentration--drop}
\end{figure}

\begin{figure}
                \centering
                \includegraphics[width=0.5\textwidth]{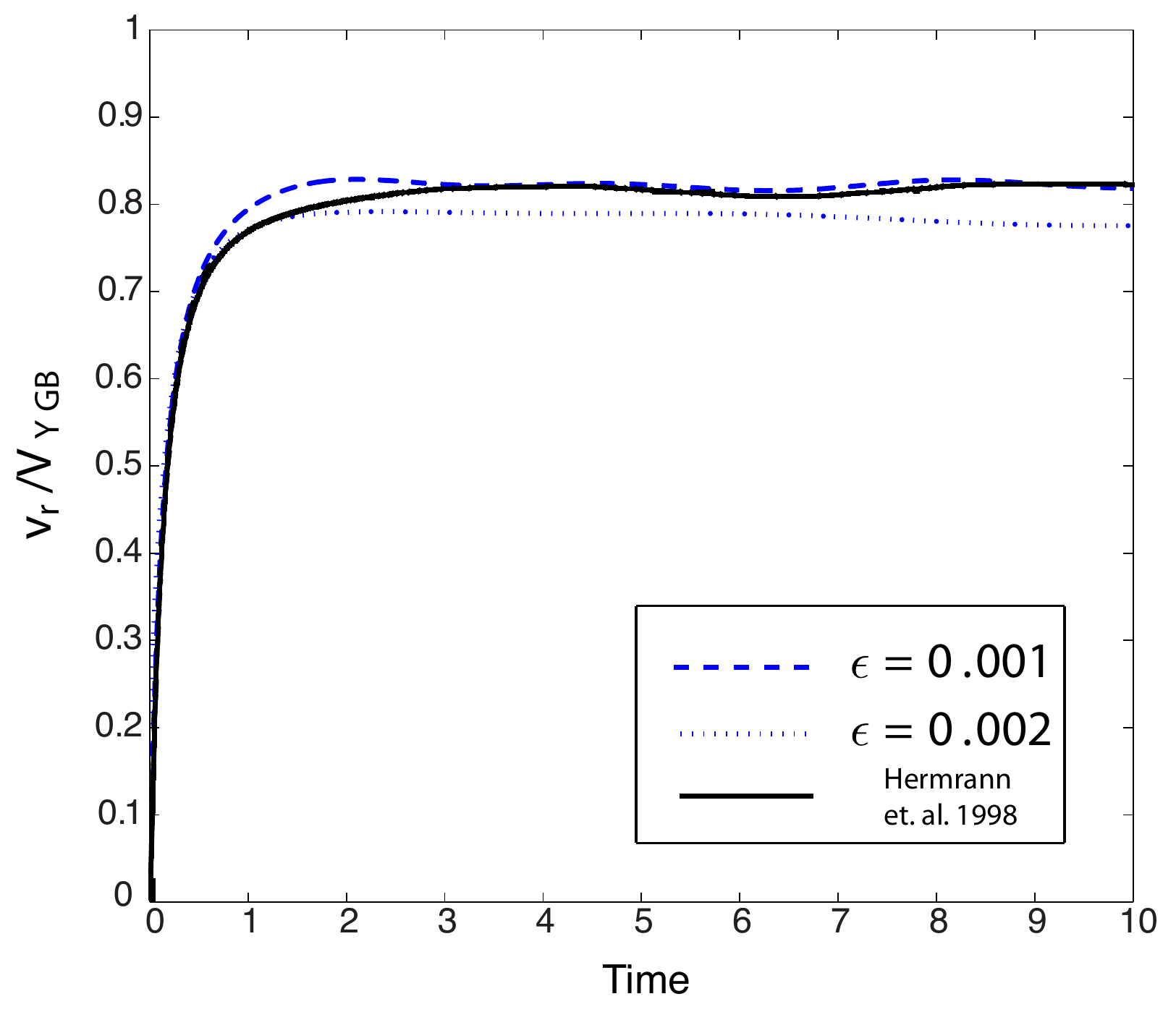}
                \caption{The time evolution of normalized migration velocity of a drop. The dashed lines are our numerical results for a 2D planar drop ($v_r$), while the solid line represents the numerical results by using level-set method.}\label{YGB--velocity}
\end{figure}

\begin{figure}
                \centering
                \includegraphics[width=0.9\textwidth]{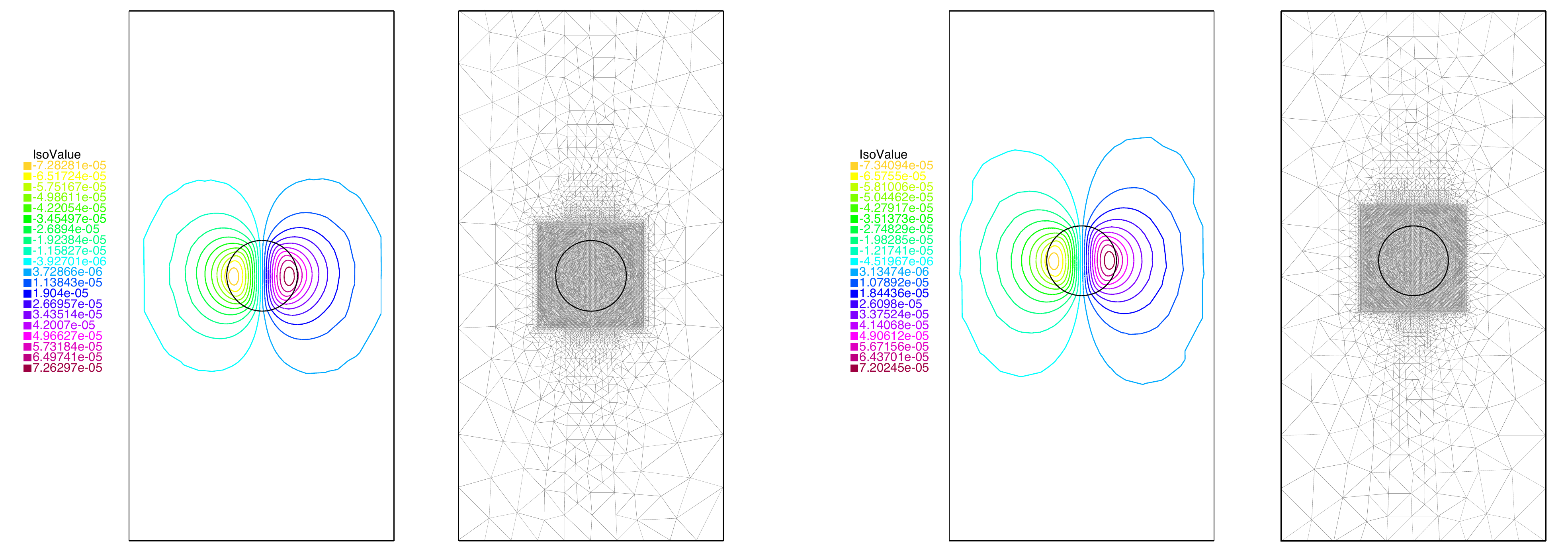}
                \caption{The drop interface (black) and the streamlines (colorful lines, left), and the meshes (grey lines, right) at $t=1$ and $t=50$. Positive values of the stream-function indicate the clockwise circulation and negative values of the stream-function indicate the counterclockwise circulation.}\label{streamline--YGB}
\end{figure}

The thermocapillary migration of a drop was first examined experimentally By Young $et$ $al.$ \cite{Young1959}, who derived an analytical expression for the terminal velocity (also known as YGB velocity) of the drop in an infinite domain. In his study, both the Marangoni and Reynolds numbers are assumed to be infinite small, such that the convective transport of momentum and energy are negligible. Instead, the terminal velocity of the drop is derived in an infinite domain with constant temperature gradient fields, and can be given in the form
\begin{eqnarray}
V_{YGB}=\frac{2U}{(2+\tilde{k})(2+3\tilde{\mu})},
\end{eqnarray}
where $U=-\sigma_TG_TR/\mu_B$ is chosen as the velocity scale, $R$ is the radius of the drop and $G_T$ stands for the constant temperature gradient, $\tilde{k}=k_A/k_B$ is the thermal conductivity ratio and $\tilde{\mu}=\mu_A/\mu_B$ is the viscosity ratio between the two fluids. In our simulation, we consider a 2D domain $\Omega$ of size $[0, 7.5R] \times [0, 15R]$ where a planar 2D circular drop of fluid A with radius $R=0.1$ is placed inside the medium of fluid B, with the drop's centre located at the centre of the box $(x_c,y_c)=(3.75R,7.5R)$. We set the initial condition for the phase field as
\begin{eqnarray}
c(x,y)=\frac{1}{2}{\rm tanh}\bigg( \frac{R-\big[(x-x_c)^2+(y-y_c)^2 \big]^{\frac{1}{2}}   }{2\sqrt{2}\epsilon}  \bigg)+\frac{1}{2}.\label{initial--c--drop1}
\end{eqnarray}
In Figure {\ref{initial--condition--for--concentration--drop}} we present the initial condition (\ref{initial--c--drop1}) for the whole domain (left hand side), and for fixed $x=3.75R$ (right hand side), where it can be observed that the area with $c=1$ represents the drop (fluid A) and the area with $c=0$ represents the medium (fluid B), between which the value of $c$ varies rapidly resulting in a diffuse interface with finite thickness. Within this transition layer,  the dotted contour line is at level $c=0.5$ representing the dividing surface $\Gamma$. No-slip boundary conditions are imposed on the top and bottom wall, and periodic boundary conditions are imposed in the horizontal direction. A linear temperature field is imposed in $y$ direction
\begin{eqnarray}
T(x,y)=T_b+\frac{T_t-T_b}{15R}y=T_b+G_T y,
\end{eqnarray}
with $T_b = 10$ on the bottom wall and $T_t = 25$ on the top wall, resulting in a constant temperature gradient, $G_T =10$. Again, we let the ratio parameter $\eta=6\sqrt{2}$ (Eq.(\ref{constant--value--eta})). Moreover, the fluid properties are shown in Table \ref{tab:kd}.
\begin{table}
\begin{center}
\begin{tabular}{llll}
$\sigma_0=5\times10^{-2}$,& $\sigma_T=1.25\times10^{-3}~~({\rm at}~~T_{ref}=T_b)$, &$\mu_A=\mu_B=0.2$,&$m_C=0.1\epsilon.$
\end{tabular}
  \caption{The physical properties of two fluids for example of thermocapillary migration (A and B stand for the fluid A and B separately).}
  \label{tab:kd}
\end{center}
\end{table}
Using these values, the theoretical terminal velocity of a spherical drop can be given as 
\begin{eqnarray}
V_{YGB} = 8.333 \times10^{-4}.\label{value--velocity}
\end{eqnarray}
Numerically, we use the following equation to calculate the rise velocity $v_r$ of the drop for our phase-field model,
\begin{eqnarray}
v_r=\frac{\int_{\Omega } c \bold{v}\cdot \hat{\bold{j}}~ {\rm d}V }{\int_{\Omega } c~{\rm d}V },\label{average--drop--velocity}
\end{eqnarray}
where $\hat{\bold{j}}$ is the component of the unit vector in $y$ direction.\\
Figure \ref{YGB--velocity} shows the temporal evolution of the drop velocity normalized by $V_{YGB}$ between two different interface capturing methods, phase-field method and level set method (\cite{Herrmann2008}). Similar to the previous example in $\S 6.3$, we compute our model by using two different interfacial thickness corresponding to $\epsilon=0.002$ and $0.001$. Both the phase-field method and level-set method seem to converge to a value of $v_r/V_{YGB} = 0.8$, roughly $20\%$ different from the theoretical prediction. The reason for this discrepancy is two-fold. For one, and most importantly, the theoretical rise velocity is for an axisymmetric sphere, whereas our simulations are carried out for a planar 2D drop. The second reason is that the simulations include small blockage effects from the finite computational domain size as well as minute deformations of the drop, whereas the theoretical formula assumes an infinite domain and a non-deformable drop. As the thickness of the diffuse interface decreases, our results seem to coverage to the that obtained by level-set method (\cite{Herrmann2008}). For the case $\epsilon=0.001$, we present the streamlines together with the moving interface at $t=1$ and $t=50$ in Figure \ref{streamline--YGB}, where we observe that the streamlines for both cases exhibit the similar patterns, with two asymmetric recirculation around the drop. Figure {\ref{streamline--YGB}} shows the meshes together with the drop interface at $t=1$ and $t=50$. Here the size of the smaller frame is set to be $[3R\times3R]$, in which we take the shortest edge of the grids inside the frame as $15R/1000=\epsilon$, so that at least 7-9 grid cell (corresponding to the definition of the interfacial thickness) is located across the interface to ensure accuracy of our computations. In addition, the moving velocity of the frame is set to be equal to the drop rising velocity $v_{frame}=v_r$, such that, through this relative long-term behavior, the rising drop is always kept inside the smaller moving frame.
\subsection{Thermocapillary migration with finite Marangoni number}
We now compute the example of the thermocapillary motion of a drop with finite Marangoni numbers. Due to the finite Marangoni numbers, the energy equation (\ref{211sp--sys--quasi--en}) is coupled with the momentum equation (\ref{211sp--sys--quasi--mom}). This is expected to result in a reduction of the tangential temperature gradients at the drop interface due to the interfacial flow driven by the Marangoni stress, which in turn will also be reduced.
In this simulation, we consider a 2D domain $\Omega$ of size $[0, 10R]\times[0, 15R]$, where a planar 2D circular drop of fluid A with radius $R=0.5$ is placed inside the medium of fluid B, with the drop's centre located at the centre of the box $(x_c,y_c)=(0, 1.5R)$. At $t =0$, Eq.(\ref{initial--c--drop1}) is employed as the initial condition for the phase variable, and a linear temperature distribution from $T_{b}=0$ at the bottom to $T_{t}=1$ at the top is imposed for the bulk liquid, and we assume that the drop has the same initial linear temperature distribution as the bulk liquid. Again, no slip boundary conditions are imposed on the top and bottom boundaries, and periodic boundary conditions are imposed in the horizontal direction. The two fluids are assumed to have the same densities and viscosities. We set the thermal conductivity $k_1=0.1$ for the drop and $k_2=1$ for the bulk fluid. In this section, the non-dimensionalized system equations (\ref{non--sys--quasi--mass})-(\ref{non--sys--quasi--en}) are computed, where we set the non-dimensional parameters as $\epsilon=0.002$, $Re=10$, $M=1$, $Pe=100/\epsilon$, $Ca=1$, $Ec=1$. Five different values of Marangoni number are employed for the computations, such that $Ma=50, 100, 500, 1000, 1500$.\\
Figure 9 shows the velocity of the drop versus time for the five cases. As the time processes, the rise velocity reduces in all five cases, where we can observe that the increase in Ma leads to the decrease in the rise velocity, which is consistent with the simulations in \cite{Herrmann2008, Yin2008, Zhao2010}.\\
Figure 10 shows snapshots of the isotherms at 4 different times for the corresponding three cases, where the dependence of the migration velocity on the Marangoni number can be easily explained. Obviously, the enhanced convective transport of momentum and heat with the increase of the Marangoni number results in more disturbances of the temperature field. Inside the drop, as we increase the Marangoni number, the larger variations can be observed, leading to a substantial reduction in the surface temperature gradient and the corresponding rise velocities.
\begin{figure}
                \centering
                \includegraphics[width=0.8\textwidth]{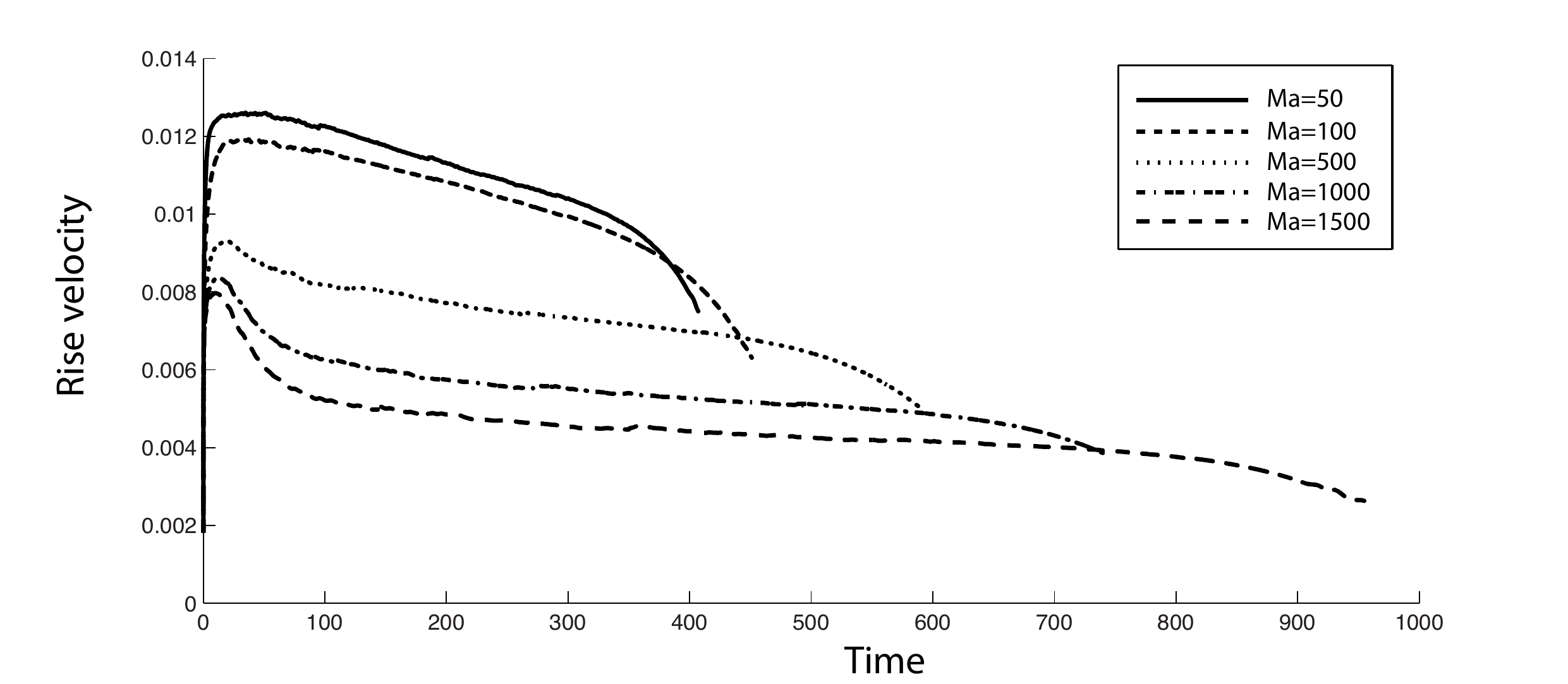}
                \caption{The time evolution of rise velocity of a drop with different finite Marangoni number.}\label{risevelocity}
\end{figure}
\begin{figure}
                \centering
                \includegraphics[width=0.9\textwidth]{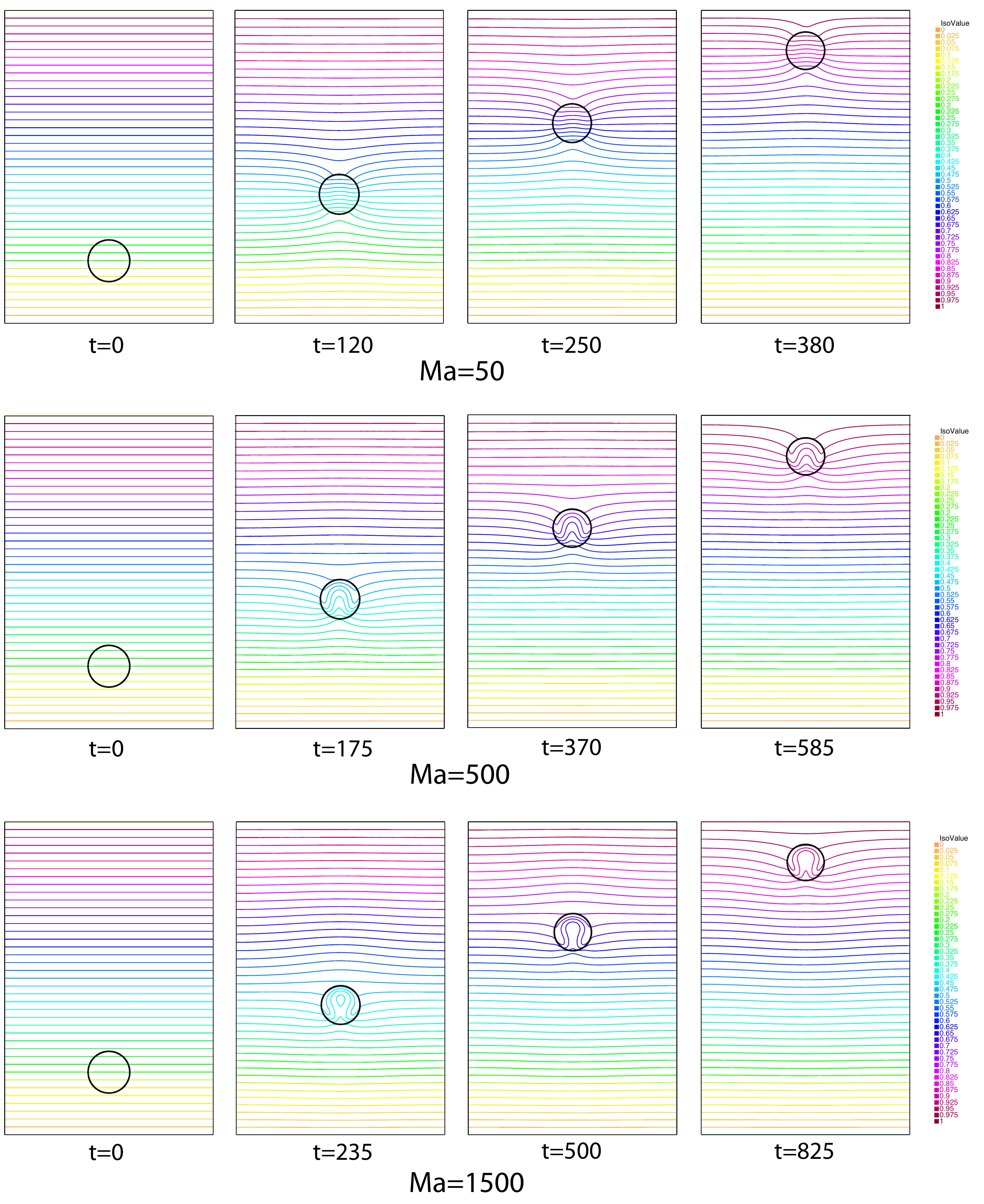}
                \caption{The snapshots of drop interface (black) and isotherms (colorful lines) for different time and different Ma as indicated.}\label{streamline--YGBq}
\end{figure}
\Section{Conclusion and future work}
In this paper, we present a thermodynamically consistent phase-field model for two-phase flows with thermocapillary effects, which allows the binary incompressible fluid (quasi-incompressible fluid) to have different physical properties for each component, including densities, viscosities and thermal conductivities. To the best of our knowledge, such a phase-field model is new. We chose the mass concentration as the phase variable, where the corresponding variable density and mass-averaged velocity lead to a quasi-incompressible formulation for the binary incompressible fluid. As the thermocapillary effects are produced by the non-homogenous distribution of a temperature dependent (linearly) surface tension, we introduce the square-gradient (Cahn-Hilliard) term into the internal energy and entropy of our phase-field model, so that the interfacial free energy that is associated with the surface tension in our model can be linearly dependent on the temperature. Our model equations, including mass balance equation, Navier-Stokes equation with extra stress term, advective Cahn-Hilliard equation, energy balance equation and entropy balance equation, are derived within a thermodynamic frame based on entropy generation. Comparing with the classical energy balance equation employed by other phase-field models, the non-classical terms associated with the square-gradient term appear in our energy balance equation (\ref{sys--quasi--en}) accounting for the energy spent by the variations of the phase field. In addition, we verify the first and second thermodynamic laws from the system of equations to show that thermodynamic consistency is maintained in our model. Moreover,  we also verified that our system equations satisfy the important modelling properties, namely the Onsager reciprocal relations and Galilean invariance.\\
In the sharp-interface analysis, we show that the system of equations and jump conditions at the interface for the classical sharp-interface model are recovered from our model, which reveals the underlying physical mechanisms of the phase-field model, and provides a validation of our model. It is worth mentioning that, in the jump condition of the momentum balance, we identify the square-gradient term of the free energy as the surface tension (Eq.(\ref{surface--term--difinition})) of our phase-field model. We further relate it to the physical surface tension through a ratio parameter, where a relation (Eq.(\ref{equ--to--confirm--coef})) is provided to determine the value of this parameter.\\
We also compute three examples, including thermocapillary convection in a two-layer fluid system and thermocapillary migration of a drop. The results for the first two examples are in good agreement with the existing analytical and numerical solutions quantitatively, which validates our phase-field model. Thus, on the whole, we conclude that the phase-field model can be very suitable for simulating multiphase flows with thermocapillary effects.\\
In the future work, besides exploring various applications and extensions of the model, we intend to provide an asymptotic analysis of the solution of the model, and use it as a further validation of our model. For the phase-field model developed here, we will present a thermodynamic consistency preserving numerical method with the corresponding numerical results in a forthcoming work \cite{Guo--new2014}.
\section*{Acknowledgement}
Z. G. was partially supported by the Chinese Scholarship Council (No.2011646021) for studying at the University of Dundee. P. L. was partially supported by the Fundamental Research Funds for Central Universities (No.06108038 and No.06108137).\bibliographystyle{plain}
\bibliography{ReferenceforPhasefield}

\end{document}